\documentclass[12pt,letter,doublespace]{article}
\usepackage{amssymb}
\usepackage{natbib,graphicx,setspace,lscape,longtable}
\usepackage{natbib,epsfig,graphicx}
\usepackage{amsmath,amsthm,amssymb}
\usepackage{color,xcolor}
\usepackage{hyperref}
\RequirePackage[mathlines, displaymath]{lineno}
\usepackage{multirow}
\bibpunct{(}{)}{;}{a}{,}{,}

\makeatletter
\def\singlespace{\def\baselinestretch{1}\@normalsize}

\def \bec{\begin{center}}
\def \enc {\end{center}}
\def \bee {\begin{eqnarray*}}
\def \ene {\end{eqnarray*}}

\def \bear{\begin{array}}
\def \enar{\end{array}}

\def \bs{\begin{slide}}
\def \es{\end{slide}}

\makeatletter
\def\singlespace{\def\baselinestretch{1}\@normalsize}

\@addtoreset{equation}{section}
\renewcommand{\theequation}{\thesection.\arabic{equation}}

\renewcommand{\baselinestretch}{1.5}

\newcommand{\bx}{{\bf x}}

\def\bec{\begin{center}}
\def\enc{\end{center}}

\numberwithin{equation}{section}  

\newtheorem{thm}{Theorem}[section]
\newtheorem{lem}{Lemma}[section]

\begin{document}

\title{A Distribution-Free Test of Independence and
Its Application to Variable Selection}

\author{Hengjian Cui \ \ and \ \ Wei Zhong\\
Capital Normal University \& Xiamen University}
\date{November 12, 2017}
\maketitle{}
\ \\

\ \\
\noindent{Running Head: A Distribution-Free Test of Independence}

\begin{footnotetext}[1]
{Hengjian Cui is Professor, Department of Statistics, Capital Normal
University, China. Email: hjcui@bnu.edu.cn. His research was
supported by National Natural Science Foundation of China (NNSFC)
grants 11071022, 11028103, 11231010, 11471223, Foundation of Beijing
Center for Mathematics and Information Interdisciplinary Sciences and
Key project of  Beijing Municipal Educational Commission KZ201410028030.
Wei Zhong is the corresponding author and Associate Professor of Wang Yanan Institute for Studies in Economics,
Department of Statistics, School of Economics and Fujian Key Laboratory of Statistical Science,
Xiamen University, Xiamen, 361005,
China.  Email: {wzhong@xmu.edu.cn.}
His research was supported by NNSFC grant 11301435, 11401497 and the Fundamental Research Funds for the
Central Universities 20720140034.
{All authors equally contribute to this paper, and the authors are listed in the alphabetic order.}
}
\end{footnotetext}

\newpage
\begin{abstract}
Motivated by the importance of measuring the association between the response and predictors in high dimensional data, we propose a new distribution-free test of independence between a categorical response variable $Y$ and a continuous predictor $X$ based on mean variance (MV) index.
The mean variance index can be considered as the weighted average of Cram\'{e}r-von Mises distances between the
conditional distribution functions of $X$ given each class of $Y$ and the unconditional distribution function of $X$.
The mean variance index is zero if and only if $X$ and $Y$ are independent.
In this paper, we propose a new MV test between $X$ and $Y$
and it enjoys several appealing merits. First, under the independence between $X$ and $Y$,
we derive an explicit form of the asymptotic null distribution, $\sum_{j=1}^{+\infty} {\chi_j^2(R-1)}/{\pi^2j^2}$, where $\chi_j^2(R-1), j=1,2,\ldots $, are independent $\chi^2$ random variables with $R-1$ degrees of freedom and $R$ is the fixed number of classes of $Y$.
It provides us with an efficient and fast way to compute the empirical p-value in practice.
Second, we can allow  $R$ diverge slowly to the infinity as the sample size increases and
the limiting null distribution of the standardized test statistic is a standard normal distribution.
Third, it is essentially a rank test and thus distribution-free.
No assumption on the distributions of two random variables is required and the test statistic is invariant under one-to-one transformations.
It is resistant to heavy-tailed distributions and extreme values in practice. We assess its excellent performance by Monte Carlo simulations. As its important application, we apply the MV test to  high dimensional colon cancer gene expression data to detect the significant genes associated with the tissue syndrome.
\end{abstract}

\


\

\newpage

\pagestyle{plain}

\section{Introduction}
One of the fundamental goals of data analysis and statistical inference is to understand the relationship among random variables.
In many scientific researches, it is of importance and interest to test whether two random variables are statistically independent of one another. Many real-life examples can be found in finance, physics, biology and medical science, etc.
For instance, the genetics researchers may be interested in testing the independence between some inherited disease and a single-nucleotide polymorphism (SNP) or whether two groups of genes are associated in high dimensional genetic data.
The medical researchers may want to understand the relationship between the lung cancer and the smoking status.

As a fundamental statistical problem, testing whether two random variables are independent or not has received much attention in the literature.
When two random variables are both categorical, the classic Pearson's chi-square test is applied to test their statistical independence.
Note that the independence of two random variables $X$ and $Y$ is equivalent to $H_0: F_{XY}=F_XF_Y$, where $F_{XY}$ denotes the joint distribution function of $(X,Y)$ and $F_X$ and $F_Y$ denote the marginal distributions of $X$ and $Y$, respectively.
\cite{Hoeffding:1948} proposed a test of independence based on the difference between the joint distribution function and the product of marginals.
The Hoeffding's test statistic is
\begin{eqnarray} \label{H}
H_n=n \int\int\left[\hat{F}_{XY}(x,y)-\hat{F}_X(x)\hat{F}_{Y}(y)\right]^2 d \hat{F}_{XY}(x,y),
\end{eqnarray}
where $\hat{F}$ denotes the empirical distribution function.
This is also the well-known Cram\'{e}r-von Mises criterion between the joint distribution function and the product of marginals.
\cite{Rosenblatt:1975} considered a measure of dependence based on the difference between the joint density function and the product of marginal densities.
To consider the quadratic distance between the joint characteristic function and the product of the marginal characteristic functions,
\cite{Szekely:Rizzo:Bakirov:2007} and \cite{Szekely:Rizzo:2009} defined a distance covariance (DC) between two random vectors $X\in\mathbb{R}^p$ and $Y\in\mathbb{R}^q$  by
\begin{eqnarray} \label{dcov}
V^2(X,Y)=\int_{\mathbb{R}^{p+q}}\left|\phi_{XY}(t,s)-\phi_{X}(t)\phi_{Y}(s)\right|^2\omega(t,s) dtds,
\end{eqnarray}
where $\phi_{XY}(t,s), \phi_{X}(t), \phi_{Y}(s)$ denote the joint characteristic function, the marginal characteristic functions of $X$ and $Y$, respectively, and  $\omega(t,s)$ is a positive weight function. $V^2(X,Y)=0$ if and only if $X$ and $Y$ are independent.
They further proposed a test of independence based on the statistic
$nV^2_n(X,Y)/S_2$, where $V^2_n(X,Y)$ is the estimator for $V^2(X,Y)$ by
using the corresponding empirical characteristic functions and $S_2=n^{-4}\sum_{k,l=1}^n|X_k-X_l|_p\sum_{k,l=1}^n|Y_k-Y_l|_q$
 in which $\{(X_i,Y_i), i=1,2,\ldots,n\}$ is a random sample of $(X,Y)$.
Under the existence of moments, it was proved that $nV^2_n(X,Y)/S_2$ converges in distribution to a quadratic form $\sum_{j=1}^{+\infty}\lambda_jZ_j^2$, where $Z_j$ are independent
standard normal random variables and the values of $\lambda_j$ depend on the distribution of $(X,Y)$.
Recently, \cite{HHG:2013} developed a consistent multivariate test of association based on ranks of distances.
\cite{Bergsma:Dassios:2014} proposed another consistent test of independence based on a sign covariance related to Kendall's tau.

In this paper, we propose a novel distribution-free test for the independence between
a categorical random variable and a continuous one based on mean variance (MV) index.
{It is important to understand the relationship between a categorical variable and a continuous one in practice, such as the relationship between the SNP and a continuous genetic trait, the tumor class and gene expression levels (continuous), or the social status and the family income, etc.}
Let $Y$ be a categorical variable with $R$ classes $\{y_1,y_2,\ldots,y_R\}$, and $X$ be a continuous variable.
The MV index can be considered as the weighted average of Cram\'{e}r-von Mises distances between the
conditional distribution functions of $X$ given each $Y=y_r$ and the unconditional distribution function of $X$. Note that
the MV index equals to 0 if and only if $X$ and $Y$ are statistically independent.
Thus, the MV index can be used to construct a test statistic for independence.
The proposed MV test enjoys several advantages. (1)
Under the null hypothesis of independence between two variables, the asymptotic null distribution has an explicit form {\color{black}when $R$ is fixed}.
That is, $\sum_{j=1}^{+\infty} {\chi_j^2(R-1)}/{\pi^2j^2}$, where $\chi_j^2(R-1), j=1,2,\ldots $, are independent
$\chi^2$ random variables with $R-1$ degrees of freedom.
It provides us with an efficient and fast way to compute the critical value and
make a test decision quickly in practice.
(2) The number of classes $R$ can be allowed to
approach infinity with the sample size $n$ at a relatively slow rate.
The limiting null distribution of the standardized MV statistic is a standard normal distribution. It is convenient to obtain any critical value in practice using an approximated normal distribution when $R$ is large.
(3) The proposed test is essentially a rank test and thus distribution-free.
Thus, the MV test statistic is invariant for any fixed $n$ under one-to-one transformations and resistent to heavy-tailed distributions and extreme values in practice. Numerical studies show that the MV test has a higher or comparable
power performance compared with the existing methods even when $X$ is generated from a standard Cauchy distribution. Furthermore, there is no distribution assumption required to derive the asymptotic null distributions. This merit is not shared by the distance covariance test \citep{Szekely:Rizzo:Bakirov:2007} whose asymptotic null distribution depends on the distribution of $(X,Y)$ and has no explicit form.

The rest of this paper is organized as follows. In Section 2, we introduce the mean variance index and its properties. Main results are included in Section 3, where we will propose a new distribution-free MV test and derive its asymptotic distributions.
In Section 4, we study the power performance of the new test compared with the existing alternative methods using  Monte Carlo simulations and a real-data application. Section 5 discusses some extensions. Technical proofs are given in the Appendix.

\section{Mean Variance Index}

In this section, we briefly introduce the mean variance index defined for a continuous random variable and
a categorical one. Let $X$ be a continuous random variable with a
support $\mathbb{R}_X$ and  $Y$ be a categorical random variable with $R$ classes $\{y_1,y_2,\ldots,y_R\}$.
The mean variance (MV) index of $X$ given $Y$ defined in \cite{Cui:Li:Zhong:2014} by
\begin{eqnarray}\label{def1}
MV(X|Y) = E_X[Var_Y(F(X| Y))],
\end{eqnarray}
where $F(x|Y)=\mathbb{P}(X\leq x|Y)$ denotes the conditional distribution function of $X$ given $Y$.
We further let $F(x)=\mathbb{P}(X\leq x)$ denote the unconditional distribution function of $X$,
and $F_r(x)=\mathbb{P}(X\leq  x| Y=y_r)$ be the conditional distribution function of $X$ given $Y=y_r$.
\cite{Cui:Li:Zhong:2014} showed that $MV(X|Y)$ can be represented as the following quadratic form between
$F(x)$ and $F_r(x)$,
\begin{eqnarray}\label{def2}
MV(X| Y)= \sum_{r=1}^R p_r\int \left[F_{r}(x) -F(x)\right]^2dF(x),
\end{eqnarray}
where $p_r=\mathbb{P}(Y=y_r)>0$ for all $r=1,\ldots, R$.
It is worth noting that $MV(X|Y)$ can be considered as the weighted average of
Cram\'{e}r-von Mises distances between the conditional distribution functions of $X$ given each $Y=y_r$ and the
unconditional distribution function of $X$.
This observation further implies the following Lemma.
{\lem\label{lem21}
 $MV(X| Y)=0$ if and only if $X$ and $Y$ are statistically independent.
}

Lemma \ref{lem21} indicates that the MV index $MV(X|Y)$ can measure any dependence
between a continuous random variable and a categorical one.
Due to this property, we will propose a test of independence between $X$ and $Y$
based on their MV index and develop the associated asymptotic  distributions in the later section.

Next, we provide a consistent estimator for $MV(X|Y)$.
Suppose that $\{(X_i, Y_i): i=1,\ldots, n\}$ with the sample size $n$
is randomly drawn from the population distribution of $(X,Y)$.
Using the idea of method of moments,
$MV(X|Y)$ can be estimated by the following statistic
\begin{eqnarray} \label{mvhat}
\widehat{MV}(X| Y) = \frac 1n \sum_{r=1}^R\sum_{i=1}^n \hat p_r\left[\hat F_{r}(X_{i}) - \hat F(X_{i})\right]^2,
\end{eqnarray}
where $\hat F(x)=n^{-1} \sum_{i=1}^nI\{X_{i}\leq x\}$ is the empirical unconditional distribution function
of $X$, $\hat F_{r}(x) = \sum_{i=1}^n I\{X_{i}\leq x, Y_i=y_r\}/\sum_{i=1}^n I\{Y_i=y_r\}$ is the
empirical conditional distribution function of $X$ given $Y=y_r$, and
$\hat p_r = n^{-1} \sum_{i=1}^n I\{Y_i=y_r\}$ denotes the sample proportion of the $r$th class,
where $I\{\cdot\}$ represents the indicator function.
The following lemma demonstrates the consistency of the proposed estimator for $MV(X|Y)$,
which is  the direct corollary of Theorem 2.1 in \cite{Cui:Li:Zhong:2014}.

{\color{black}
{\lem\label{lem22}
 Suppose $R=R(n)=O(n^{\kappa})$ for some $0\le \kappa<1$ and there exist two positive constants $c_1$ and $c_2$ such that
$\displaystyle c_1/R \leq \underset{1\leq r\leq R}{\min}p_r \leq \underset{1\leq r\leq R}{\max}p_r \leq c_2/R $.
Then, for any  $\epsilon \in (0,1/2)$, there exists a positive constant $c>0$ such that
\begin{eqnarray}~~~~
\mathbb{P}\left\{\left|\widehat{MV}(X| Y) - MV(X|Y)\right| \geq \epsilon \right\}
\leq O(n)R\exp\left\{- \frac{cn}{R}\epsilon^2\right\}\rightarrow 0,
\end{eqnarray}
as $n\rightarrow \infty$. That is,
$\widehat{MV}(X| Y) \overset{p}{\rightarrow} MV(X|Y)$, as $n\rightarrow \infty$.
Hence, $\widehat{MV}(X| Y)$ is consistent to the mean variance index $MV(X|Y)$.
}
\vspace{.2cm}

\noindent {\sc Remark:} The condition $\displaystyle c_1/R \leq \underset{1\leq r\leq R}{\min}p_r \leq \underset{1\leq r\leq R}{\max}p_r \leq c_2/R $ requires that the proportion of each class of $Y$ cannot be either too small or too large as $n$ increases. Here, $R=O(n^{\kappa})$ is allowed to be diverging at a relatively slow rate of the sample size $n$.
If $R$ is fixed when $\kappa=0$, the condition $\displaystyle c_1/R \leq \underset{1\leq r\leq R}{\min}p_r \leq \underset{1\leq r\leq R}{\max}p_r \leq c_2/R $
is automatically satisfied and the result also holds.
}

\section{Main Results}
\subsection{Mean Variance Test of Independence}

\noindent In this section, we will present a distribution-free test of independence
between a continuous random variable $X$ and a categorical one $Y$
based on their mean variance index. We consider the following testing hypothesis:
\begin{eqnarray*}
&& H_0: \mbox{$X$ and $Y$ are statistically independent.}\\
&\mbox{versus}& H_1: \mbox{$X$ and $Y$ are not statistically independent.}
\end{eqnarray*}
Note that the null hypothesis is equivalent to that
the conditional distribution function of $X$ given $Y=y_r$ is
always equal to the unconditional distribution function $X$ for any $r=1,\ldots, R$.
That is, $F_{r}(x)=F(x)$. Thus, the previous hypothesis can be rewritten as
\begin{eqnarray*}
&& H_0: F_{r}(x)=F(x) \mbox{~for any~} x \mbox{~and~} r=1,\ldots, R.\\
&\mbox{versus}& H_1: F_{r}(x)\ne F(x) \mbox{~for some~} x \mbox{~and~} r=1,\ldots, R.
\end{eqnarray*}
To test $H_0$, we naturally consider the difference between
each $F_{r}(x)$ and $F(x)$. Note that the proposed MV index (\ref{def2})
is the weighted quadratic distance between $F_{r}(x)$'s and $F(x)$
with the proportion of each class as weights.
Therefore, we propose a new test statistic based on the sample-level MV index
\begin{eqnarray}\label{mvt}
T_n=n\widehat{MV}(X| Y) = \sum_{r=1}^R\sum_{i=1}^n \hat p_r\left [\hat F_{r}(X_{i}) - \hat F(X_{i})\right]^2.
\end{eqnarray}
The larger value of $T_n$ provides a stronger evidence against the null hypothesis $H_0$.
We name the new test as the Mean Variance (MV) Test of independence.

\begin{figure}[ht]
\centering
\includegraphics[scale=0.58]{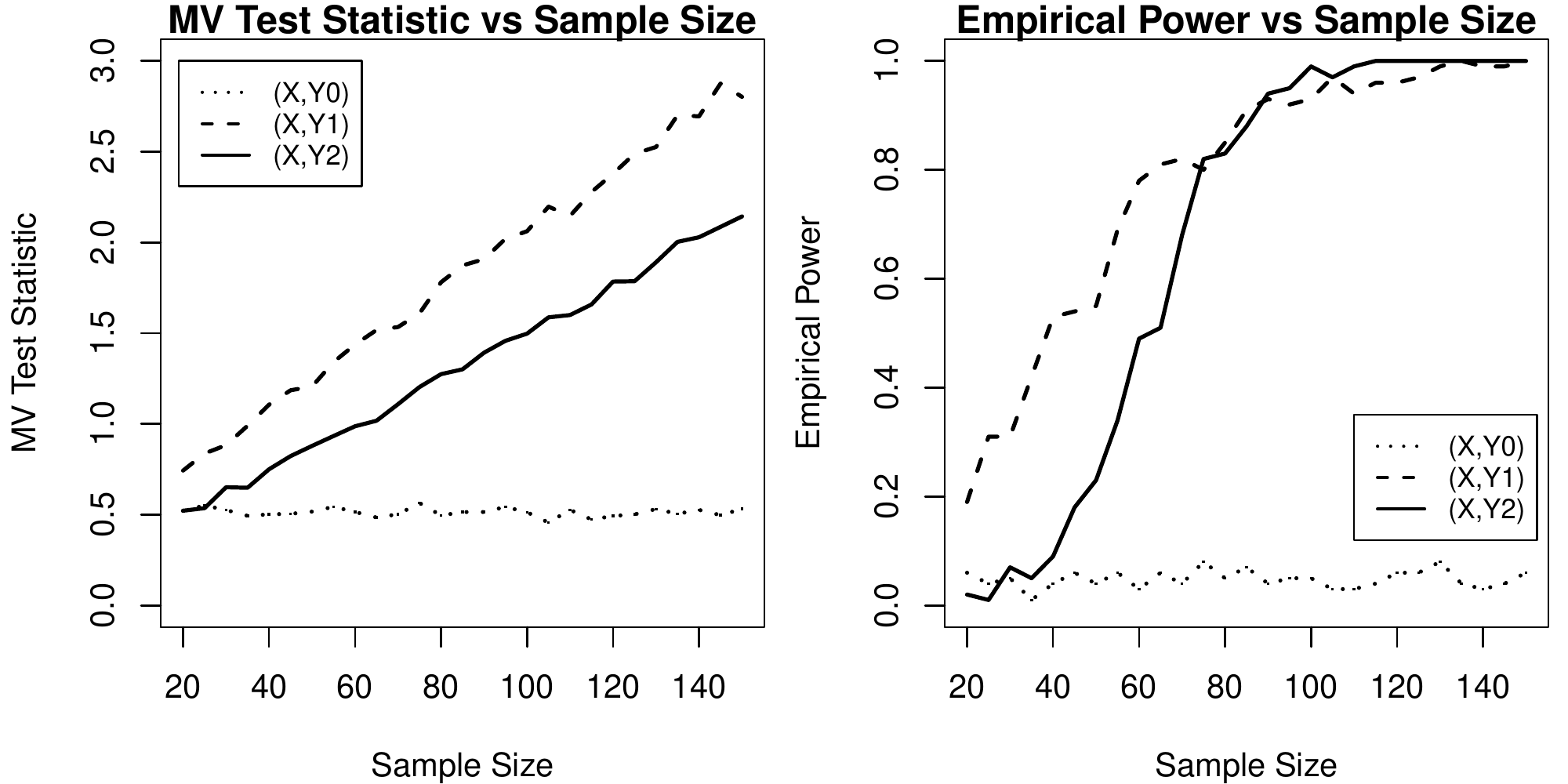}
\caption[]{\label{fig1} \small \it The left panel depicts the MV test statistic values against the sample sizes and the right panel displays the empirical powers of MV test of independence against the sample sizes. In the both panels, the dotted line, the dashed line and the solid line denote the tests of independence between $(X,Y_0)$, $(X,Y_1)$ and $(X,Y_2)$, respectively.}
\end{figure}

Before studying its theoretical properties of the MV test, we run a simple simulation example to get a first insight into how it performs. Let us generate a random variable $X$ from a standard normal distribution and
random variables $Z_k$ with $k=0,1,2$ by $Z_0=\varepsilon$, $Z_1=X+\varepsilon$ and $Z_2=2X^2+\varepsilon,$
where $\varepsilon \sim N(0,1)$ independent of $X$.
For each $k=0,1,2$, let $Y_k=I(Z_k\leq q_{k1})+2I(q_{k1}<Z_k\leq q_{k2})+3I(q_{k2}<Z_k\leq q_{k3})+4I(Z_k> q_{k3}),$
where $\{q_{k1}, q_{k2}, q_{k3}\}$ are the first, second and third quartiles of $Z_k$, respectively.
Thus, $Y_0$ is statistically independent of $X$ while $Y_1$ and $Y_2$ respectively depend on $X$ through a linear term
and a quadratic term, respectively. We consider the sample sizes $n$ from 20 to 150. 
For a given sample size, $T_n$ is computed for each pair of $(X,Y_k)$ and the associated p-value
is also calculated using its limiting null distribution which will be given in (\ref{nulldist}).
We conduct this simulation 100 times to compute the empirical powers or type-I error rates (if $H_0$ is true) at the nominal significance level 0.05.
The left panel of Figure \ref{fig1} depicts the mean of MV test statistic values against the sample sizes.
When $X$ and $Y_0$ are independent, the values of $T_n$ are close to zero for all of the sample sizes.
However, the  values of $T_n$ increase substantially as the sample sizes increase when $Y_k$ is dependent on $X$, for $k=1,2$.
The right panel displays the empirical powers of MV test of independence against the sample sizes.
When $H_0$ is true, i.e. $X$ and $Y_0$ are independent, the dotted line shows the empirical type-I error rates for different samples sizes.
The MV test performs well because the empirical type-I error rates are close to $0.05$ and have mean 0.048 and standard deviation 0.016. When $H_0$ is false, the empirical powers increase quickly to 1 as the sample size increases.
It indicates that the MV test is  useful  against both linear and quadratic dependence alternatives between a categorical random variable and a continuous one. More numerical studies can be seen in Section 4.

\subsection{Asymptotic Distributions of MV Test Statistic}

\noindent As aforementioned, the MV test statistic $T_n$ has a simple form and is easy to calculate and interpret.
However, it is by no means straightforward to derive its asymptotic distributions.
In this subsection, we will study the asymptotic distributions of $T_n$ with the aid of the empirical processes  theory.

First of all, we derive the asymptotic null distribution of $T_n$ when the class number $R$ is fixed.
The proof is given in the Appendix.

{\thm \label{thm:main1}  Suppose $X$ is a continuous random variable and $Y$ is a categorical random variable with a fixed number $R$ of classes. Under $H_0$,
\begin{eqnarray}\label{nulldist}
T_n=n\widehat{MV}(X| Y)  \overset{d}{\longrightarrow} \sum_{j=1}^{+\infty} \frac {\chi_j^2(R-1)}{\pi^2j^2},
\end{eqnarray}
 where $\chi_j^2(R-1)$'s, $j=1,2, \cdots$, are independently and identically distributed (i.i.d.) $\chi^2$ random variables with
$R-1$ degrees of freedom, and $\overset{d}{\longrightarrow}$ denotes the convergence in distribution.
}

Theorem \ref{thm:main1} demonstrates the appealing advantages of the proposed MV test.
First of all, under the independence between $X$ and $Y$, the asymptotic null distribution has an explicit quadratic form
$\sum_{j=1}^{+\infty}  {\chi_j^2(R-1)}/{\pi^2j^2},$ which
 provides us with an efficient way to compute the empirical p-value and draw a test conclusion in practice.
 It is very helpful especially when both the number of tests to conduct and the sample size are very large.
{Second, the MV test is essentially a rank test and thus distribution-free because the test statistic
is only based on the empirical distribution functions. There is no assumption on the distribution of $X$ or $Y$ required to prove Theorem \ref{thm:main1}
and the MV test statistic is invariant under any one-to-one transformation.}
This merit makes the MV test have a wide range of applications.
The distance covariance test does not share this feature because its asymptotic null distribution  depends on the distribution of $(X,Y)$.

{\color{black}\noindent {\sc Remark:} This theoretical result is
related to the asymptotic null distributions of some tests in the literature.
\cite{Szekely:Rizzo:Bakirov:2007} proved that the asymptotic null distribution of their distance covariance test statistic also has a quadratic form $\sum_{j=1}^{+\infty}\lambda_jZ_j^2$ where $Z_j$'s are independent
standard normal random variables, but the values of $\lambda_j$ are unknown.
Remark that, without the explicit null distribution, one has to use the permutation test to find p-value in practice,
which is computationally inefficient when the sample size or the number of tests is very large.
}

\begin{figure}[!h]
\begin{center}
\begin{minipage}[c]{1\textwidth}
\centering
\includegraphics[scale=0.61]{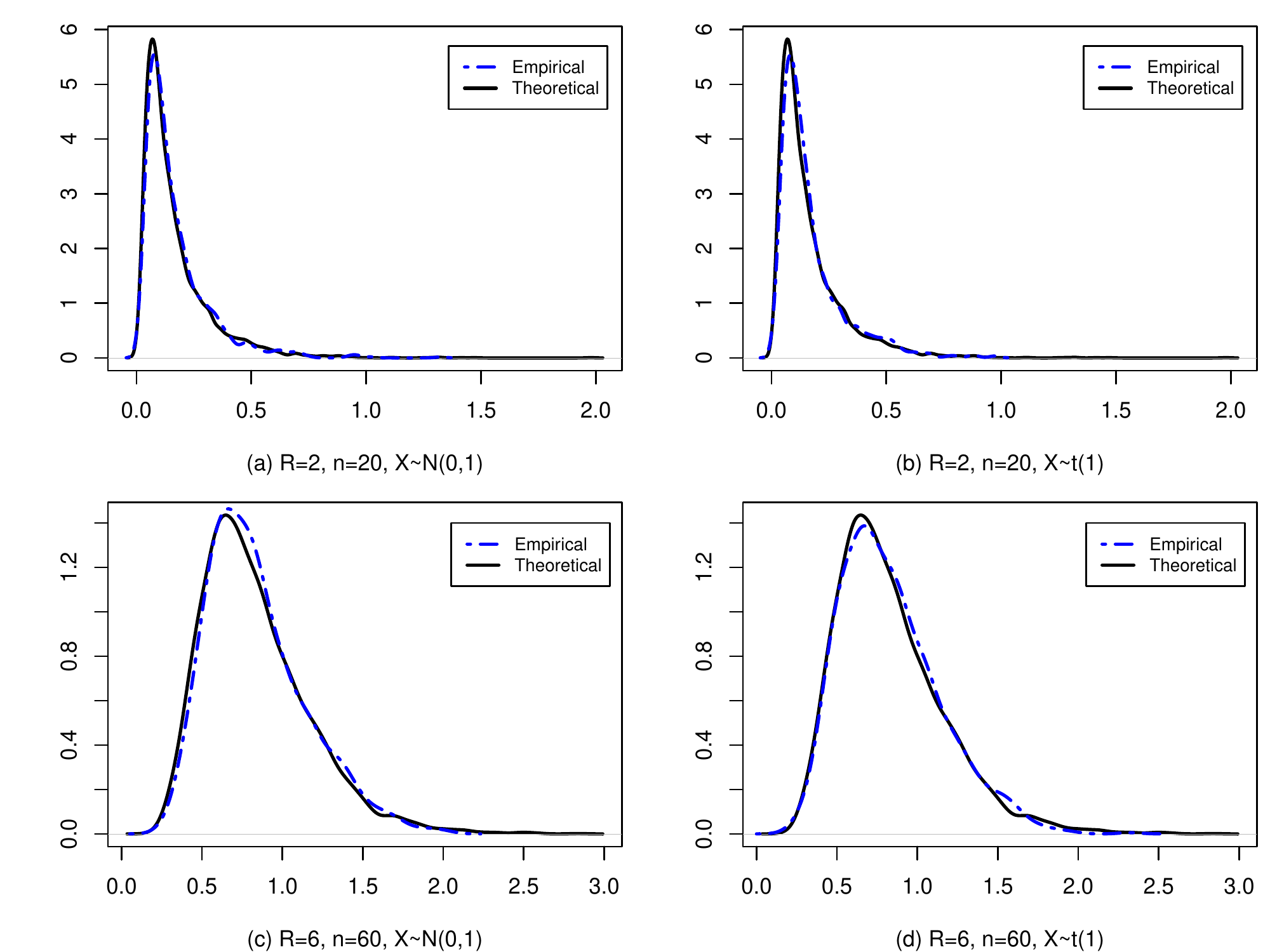}
\caption[]{\label{fig2} \small \it Comparing the empirical distribution of the MV test statistic with the asymptotic theoretical distribution under the null hypothesis. {\color{black}The empirical null distribution (broken) is a kernel density estimate using gaussian kernels based on 1000 values of $T_n$
and the asymptotic null distribution (solid) is obtained in (\ref{nulldist}).}}
\end{minipage}
\end{center}
\end{figure}

To check the validity of the asymptotic null distribution of $T_n$ obtained in Theorem \ref{thm:main1},
we compare the empirical null distribution with the asymptotic null distribution using simple simulation examples.
We generate $Y$ from a discrete uniform distribution with $R$ categories
and $X$ independently from $N(0,1)$ or $t(1)$. Note that $t(1)$  is heavily-tailed and
easy to generate extreme values. We consider
four different scenarios: (a) $R=2, n=20, X\sim N(0,1)$; (b) $R=2, n=20, X\sim t(1)$;
(c) $R=6, n=60, X\sim N(0,1)$; (d) $R=6, n=60, X\sim t(1)$.
For each scenario, we run the simulation 1000 times to obtain 1000 values of the MV test statistic $T_n$
and then compare the empirical distributions of $T_n$ with its asymptotic null distributions (see Figure \ref{fig2}).
Remark that we will elaborate how to plot the asymptotic null distribution in the next subsection.
In each panel, the two density curves are very consistent with each other, which strongly
suggests that the asymptotic null distribution in Theorem \ref{thm:main1} provides a satisfactory approximation of the null distribution
even when the sample size is relatively small.
It is worth noting that Panel (b) and (d) further show that the MV test is robust and has a reliable performance when
the distribution of $X$ is heavy-tailed and the data contain extreme values.

The next theorem shows that under the alternative hypothesis,
the MV test statistic diverges to infinity as $n\rightarrow \infty$.
In other words, if $X$ and $Y$ are dependent, i.e. $MV(X| Y) >0$, the power of the MV test to reject the false null hypothesis
converges to one as $n$ approaches the infinity. Thus, the MV test is a consistent test.

{\thm \label{thm:main2}  Suppose that the conditions assumed in Lemma 2.2 hold. Under the alternative hypothesis $H_1$, we have
\begin{eqnarray}\label{alterdist}
T_n=n\widehat{MV}(X| Y)  \overset{p}{\longrightarrow} \infty, \mbox{~~as~~} n\to \infty,
\end{eqnarray}
 where $\overset{p}{\longrightarrow}$ denotes the convergence in probability.
}

Then, we study the asymptotic normality of $\widehat {MV}(X|Y)$ which helps us to find
an expression of the asymptotic power function of the MV test.

{\thm \label{thm:main3}
Under the alternative hypothesis $H_1$, i.e. $MV(X|Y) > 0$,
we have
\begin{eqnarray}\label{alternorm}
 \sqrt {n} ( \widehat {MV}(X|Y) - MV(X|Y) ) \overset{d}{\longrightarrow} N(0, \sigma^2),
 \end{eqnarray}
where $\sigma^2= Var[\sum_{r=1}^RI_{4r}(X,Y)]$, {\color{black}where $I_{4r}(X,Y)$ is given in the Appendix.}
}

Based on Theorem \ref{thm:main3}, we can derive the following asymptotic power function of the MV test.
\begin{eqnarray} \nonumber
\beta_n(\Delta)&=&P\left(T_n>c_{\alpha}|MV(X|Y)=\Delta>0\right)\\ \nonumber
&=&P\left(\frac{\sqrt {n}(\widehat {MV}(X|Y) -\Delta)}{\sigma}> \frac{c_{\alpha}-n\Delta}{\sqrt{n}\sigma}\Big|MV(X|Y)=\Delta>0\right)\\ \label{power}
&\approx&1-\Phi\left(\frac{c_{\alpha}-n\Delta}{\sqrt{n}\sigma}\right),
\end{eqnarray}
where $\Phi(\cdot)$ is the cumulative distribution function of the standard normal distribution and
$c_{\alpha}$ denotes the $\alpha$ upper-tailed value of the asymptotic null distribution of the MV test statistic under $H_0$.
It can be observed that the power $\beta_n(\Delta)$ increases for fixed $\Delta$ and $\alpha$ as the sample size increases.
This result will also be confirmed by Monte Carlo studies in Section 4.

\subsection{Implementation of MV Test}

\begin{figure}[!h]
\begin{center}
\begin{minipage}[c]{1\textwidth}
\centering
\includegraphics[scale=0.57]{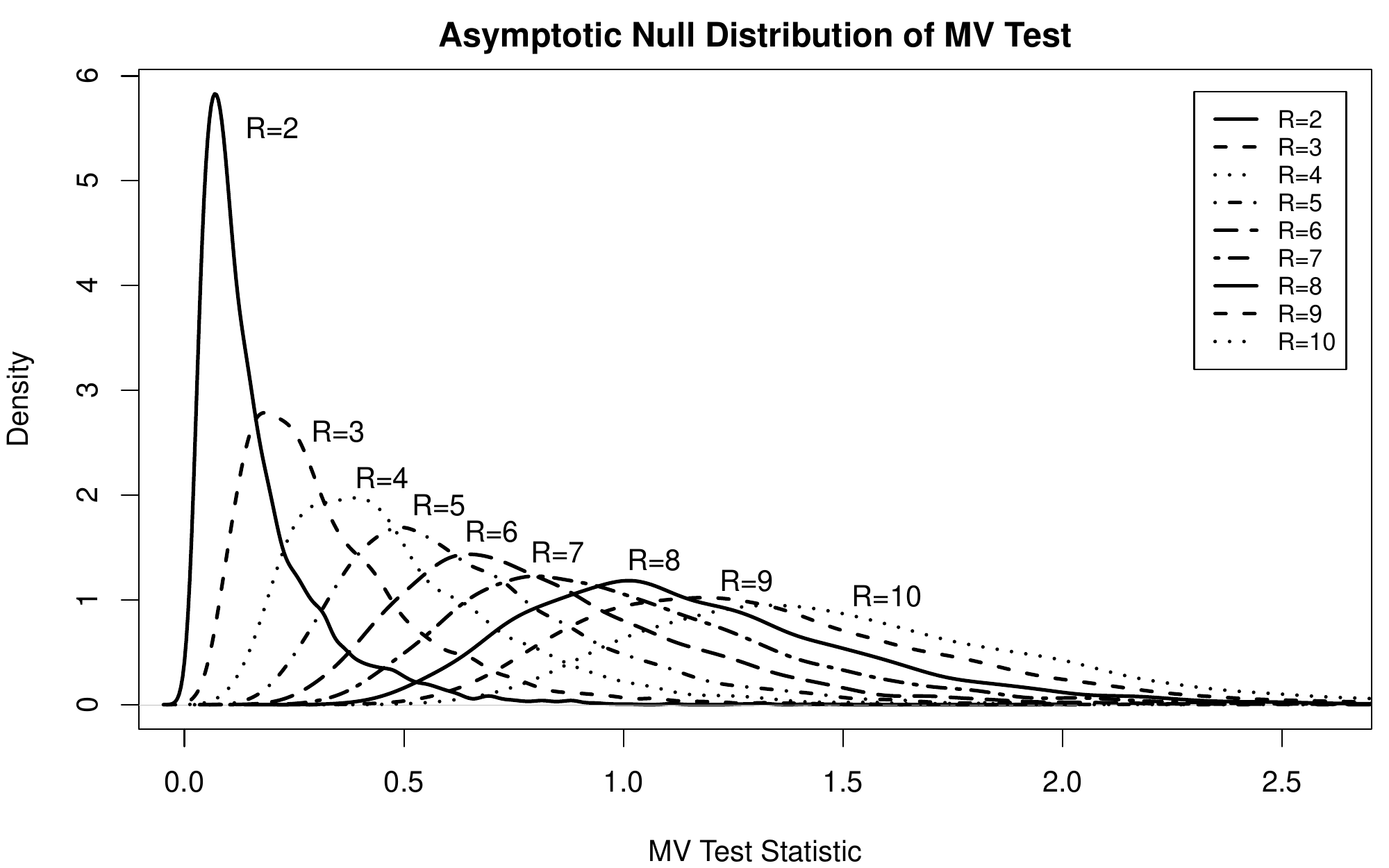}
\caption[]{\label{fig3} \small \it The asymptotic null distributions of the MV test statistic.}
\end{minipage}
\end{center}
\end{figure}

\noindent  In this subsection, we discuss the implementation of the MV test in practice.
The appealing feature of the MV test is that Theorem \ref{thm:main1}  provides
the explicit asymptotic null distribution of $T_n$ when $R$ is fixed.
Note that ${\chi_j^2(R-1)}/{\pi^2j^2}$ is ignorable when $j$ is very large.
We approximate the asymptotic null distribution  $\sum_{j=1}^{+\infty}  {\chi_j^2(R-1)}/{\pi^2j^2}$
 by $\sum_{j=1}^{N}  {\chi_j^2(R-1)}/{\pi^2j^2}$
for $N$ sufficiently large in practice.
We display the asymptotic null distributions of $R-1$ degrees of freedom, $R=2,3,\ldots, 10$, in Figure \ref{fig3}.
The density curves show that the asymptotic null distribution for each $R$ is right-skewed like a $\chi^2$ distribution and
approaches to a normal distribution as $R$ increases.

Our empirical studies show that the asymptotic null distribution performs well even when
the sample size is not large. However, if the sample size is very small,
the permutation test can be used to find the p-value for the MV test.
The permutation test is computationally efficient when the sample size is small. For example,
\cite{Szekely:Rizzo:Bakirov:2007} applied the permutation test to their distance covariance test of independence.
\cite{HHG:2013} also used the permutation samples to compute the p-value for their test of association based
on ranks of distances.
However,  the permutation test is not computationally efficient
especially when there are many pairs of random variables needed to test.
In this case, it is appealing to use our MV test based on
the explicit asymptotic null distribution to save computational complexity.

\subsection{Asymptotic Distribution when $R$ is Diverging}

\noindent The asymptotic distributions of MV test statistic have been studied before when the number of classes is fixed.
Next, we will derive its asymptotic null distribution when $R$ tends to the infinity.

{\thm \label{thm:main4}
If  $ \sqrt{R} /\underset{1\le r\le R}{\min} p_r
=o(\sqrt{n}) $ and $R\to +\infty $ as $n\to \infty$, then under $H_0$, we have
\begin{eqnarray} \label{nulldist2}
 \frac {T_n - (R-1)/6}{\sqrt{(R-1)/45}} \overset{d}{\longrightarrow}  N(0,1), \mbox{~as~} n\to \infty.
\end{eqnarray}
}

If $\underset{1\le r\le R}{\min} p_r=O(n^{-\gamma})$ where $0<\gamma<1/2$, then we can derive that $R=O(n^{\kappa})$ for some $0<\kappa<1-2\gamma$.
That is, we can allow the number of subgroups  go to the infinity with the sample size $n$ at a relatively slow rate.
This result is another distinguished merit of our test from the existing methods.
Theorem \ref{thm:main4} shows that the limiting null distribution of the MV test can be approximated by
a normal distribution with mean $(R-1)/6$ and variance $(R-1)/45$ when  $R$ is large.
To connect it to the asymptotic null distribution when $R$ is fixed in Theorem \ref{thm:main1},
one can note that  the mean and variance of  $\sum_{j=1}^{+\infty} {\chi_j^2(R-1)}/{\pi^2j^2}$ are given by
\begin{eqnarray} \label{meanvar}
~~~~~~~~~~E\left(\sum_{j=1}^{+\infty} \frac {\chi_j^2(R-1)}{\pi^2j^2}\right) = \frac{R-1}{6},~~
Var\left(\sum_{j=1}^{+\infty} \frac {\chi_j^2(R-1)}{\pi^2j^2}\right) = \frac{R-1}{45}.
\end{eqnarray}

\section{Numerical Studies}
\subsection{Monte Carlo Simulations}

\noindent
In this section, we assess the finite-sample performance of the MV test (MV)
of independence by comparing with other existing tests:
the classic Pearson's chi-square test (CS), the distance covariance test (DC) in \cite{Szekely:Rizzo:Bakirov:2007},  and
the test based on ranks of distances (HHG) in \cite{HHG:2013} in various simulation examples.
Because the Pearson's chi-square test of independence is only applicable for two discrete/categorical random variables,
we discretize equally the continuous variable into a discrete one with the same number of classes
as the categorical one. 
The permutation test with the permutated times $K=200$ is used for the DC and HHG tests since their explicit asymptotic null distributions are not available.
{\color{black} The DC and HHG tests are applied by calling the functions \emph{dcov.test} in the R package \emph{energy} \citep{Rizzo:Szekely:2014}
and
 \emph{hhg.test} in the R package \emph{HHG} \citep{Kaufman:2014}, respectively.}
Note that it is meaningless to directly apply the DC test to a categorical variable. Thus, we transfer the categorical variable
with $R$ classes to a vector of $R-1$ dummy binary variables and apply \emph{dcov.test} to this random vector instead of the original variable. {\color{black}For the MV test, we consider two ways to compute the p-value: the permutation test with $K=200$ (denoted by MV1)
and the asymptotic null distribution in (\ref{nulldist}) (denoted by MV2) for the first three examples. In Example 4, the p-value for the MV test is obtained using an approximated normal distribution  based on the asymptotic results in Theorem \ref{thm:main4}.}
All numerical studies are conducted using R code.

\textbf{Example 1.}
We randomly generate a continuous random variable $X$ from $N(0,1)$ or t(1) and independently
generate  a categorical random variable $Y$ from a discrete uniform distribution with $R$ classes.
Then, we test the independence between two random variables when $R=2$ or $6$.
The sample sizes $n$ are chosen to be 50, 75, 100, 125 and 150.
We run each simulation 1000 times to compute the empirical  type-I error rates at the nominal significance level $\alpha=0.1$
and summarize the results in Table \ref{table.ex1}.
Most of tests perform well since the empirical type-I error rates are close to the nominal significance level.
However, when $R=2$, the Pearson's chi-square test (CS) is relatively conservative because the information of $X$ may loss
substantially after discretized into a binary variable.
When $X\sim t(1)$, the distance covariance test (DC) seems conservative  due to extreme values.

\begin{table}[ht]
\begin{center}
\caption{\label{table.ex1} \it Empirical type-I error rates at the significance level 0.1 in Example 1.}
\scalebox{1}{%
\begin{tabular}{cr|ccccc|ccccc}
  \hline
 &  &  \multicolumn{5}{c|}{$X\sim N(0,1)$} & \multicolumn{5}{c}{$X\sim t(1)$} \\ \hline
  $R$     & $n$ &   MV1 &   MV2 &   DC  &    CS &   HHG  &    MV1 &   MV2 &    DC &    CS &   HHG \\  \hline
          & 50 &  0.091 & 0.091 & 0.080 & 0.055 & 0.086  &  0.091 & 0.093 & 0.079 & 0.061 & 0.089 \\
          & 75 &  0.115 & 0.118 & 0.105 & 0.065 & 0.103  &  0.100 & 0.106 & 0.095 & 0.075 & 0.102 \\
$2$      & 100 &  0.099 & 0.098 & 0.095 & 0.053 & 0.075  &  0.098 & 0.101 & 0.106 & 0.054 & 0.086 \\
         & 125 &  0.116 & 0.123 & 0.113 & 0.078 & 0.111  &  0.096 & 0.098 & 0.092 & 0.076 & 0.097 \\
         & 150 &  0.095 & 0.097 & 0.099 & 0.061 & 0.090  &  0.097 & 0.098 & 0.100 & 0.068 & 0.109 \\      \hline
          & 50 &  0.115 & 0.109 & 0.106 & 0.094 & 0.104  &  0.105 & 0.106 & 0.079 & 0.096 & 0.102 \\
          & 75 &  0.100 & 0.106 & 0.104 & 0.102 & 0.103  &  0.115 & 0.113 & 0.108 & 0.096 & 0.097 \\
$6$      & 100 &  0.093 & 0.093 & 0.094 & 0.107 & 0.083  &  0.083 & 0.091 & 0.084 & 0.092 & 0.089 \\
         & 125 &  0.109 & 0.100 & 0.100 & 0.102 & 0.097  &  0.099 & 0.105 & 0.095 & 0.112 & 0.101 \\
         & 150 &  0.105 & 0.109 & 0.098 & 0.107 & 0.104  &  0.105 & 0.107 & 0.104 & 0.103 & 0.112 \\
   \hline
\end{tabular}
}
\end{center}
\end{table}

\textbf{Example 2.}
We first randomly generate a categorical random variable $Y$ from $R$ classes $\{1,2,\ldots,R\}$
with the unbalanced proportions $p_r=P(Y=r)=2[1+(r-1)/(R-1)]/3R$, $r=1,2,\ldots,R$,
where $\{p_1,\ldots,p_r\}$ is an arithmetic progression with
$\underset{1\leq r\leq R}{\max}p_r=2\underset{1\leq r\leq R}{\min}p_r$.
For instance, when $Y$ is binary, $p_1=1/3$ and $p_2=2/3$.
Given $Y_i=r$, the $i$th predictor $X_i$ is then generated by letting $X_i=\mu_r+\varepsilon_i$, where $r=1,2,\ldots,R$.
We consider the following two choices of $R$:
(1) $R=2$, $\mu=(\mu_1,\mu_2)=(1,2)$ and $\varepsilon\sim N(0,1)$ or $t(1)$.
(2) $R=6$, $\mu=(\mu_1,\mu_2,\ldots,\mu_6)=(6,3,4,1,5,2)/3 $ and $\varepsilon\sim N(0,1)$ or $t(1)$.
In both cases, $X$ is dependent on the categories of $Y$, so the null hypothesis is false.
Table \ref{table.ex21} shows the empirical powers of each test for different sample sizes
based on 500 simulations at $\alpha=0.05$. When $X$ is normal,
all tests perform well and the MV test is slightly better than others.
When the data contain extreme values, the empirical powers of the DC, CS and HHG tests deteriorate quickly
while the MV test reasonably well.

\begin{table}[ht]
\centering
\caption{\label{table.ex21} \it Empirical powers at $\alpha=0.05$ against the sample sizes in Example 2.}
\scalebox{1}{%
\begin{tabular}{cc|ccccc|ccccc}
  \hline
   &  &  \multicolumn{5}{c|}{$X\sim N(0,1)$} & \multicolumn{5}{c}{$X\sim t(1)$} \\ \hline
 $R$ & $n$ & MV1 & MV2 & DC & CS & HHG   & MV1 & MV2 & DC & CS & HHG  \\
  \hline
   &  50 & 0.850 & 0.822 & 0.848 & 0.580 & 0.718       &  0.474 & 0.484 & 0.224 & 0.372 & 0.396   \\
   &  75 & 0.960 & 0.966 & 0.966 & 0.822 & 0.906       &  0.630 & 0.618 & 0.286 & 0.550 & 0.594     \\
$2$& 100 & 0.986 & 0.988 & 0.988 & 0.942 & 0.970       &  0.758 & 0.758 & 0.406 & 0.704 & 0.708   \\
   & 125 & 1.000 & 1.000 & 1.000 & 0.978 & 0.998       &  0.860 & 0.864 & 0.474 & 0.818 & 0.816     \\
   & 150 & 1.000 & 1.000 & 1.000 & 0.992 & 0.998       &  0.922 & 0.924 & 0.556 & 0.896 & 0.882   \\ \hline
   &  50 & 0.746 & 0.742 & 0.708 & 0.376 & 0.348       &  0.348 & 0.254 & 0.168 & 0.218 & 0.190 \\
   &  75 & 0.958 & 0.930 & 0.936 & 0.780 & 0.700       &  0.542 & 0.402 & 0.264 & 0.384 & 0.296  \\
$6$& 100 & 0.992 & 0.982 & 0.982 & 0.934 & 0.878       &  0.700 & 0.576 & 0.322 & 0.498 & 0.386  \\
   & 125 & 0.998 & 0.998 & 0.998 & 0.976 & 0.944       &  0.830 & 0.714 & 0.382 & 0.596 & 0.524   \\
   & 150 & 1.000 & 0.998 & 1.000 & 0.994 & 0.988       &  0.892 & 0.816 & 0.456 & 0.736 & 0.652   \\
   \hline
\end{tabular}}
\end{table}

Then, we consider local power analysis of all tests under contiguous sequence of alternative hypotheses.
We fix $n=100$ and consider two cases:
(1) $R=2$, $\mu=(\mu_1,\mu_2)=c(1,2)$, $\varepsilon\sim N(0,1)$ or $t(1)$;
(2) $R=6$, $\mu=(\mu_1,\mu_2,\ldots,\mu_6)=c(6,3,4,1,5,2) $ and $\varepsilon\sim N(0,1)$ or $t(1)$.
The values of $c$ vary from 0 to 1, which control the signal strength against alternatives.
When $c=0$, $X$ and $Y$ are statistically independent and $H_0$ is true;
otherwise, $H_0$ is false.
We display the empirical powers of all tests against the values of $c$ in Table \ref{table.ex22}.
The MV test has the excellent power performance in most settings especially when $X$ follows $t(1)$.

\begin{table}[ht]
\centering
\caption{\label{table.ex22} \it Empirical powers at $\alpha=0.05$ against the signal strength in Example 2.}
\scalebox{1}{%
\begin{tabular}{cr|ccccc|ccccc}
  \hline
   &  &  \multicolumn{5}{c|}{$X\sim N(0,1)$} & \multicolumn{5}{c}{$X\sim t(1)$} \\ \hline
 $R$ & c & MV1 & MV2 & DC & CS & HHG   & MV1 & MV2 & DC & CS & HHG  \\
  \hline
   & 0.0 & 0.054 & 0.056 & 0.040 & 0.022 & 0.046 &  0.060 & 0.054 & 0.050 & 0.040 & 0.044    \\
   & 0.2 & 0.156 & 0.164 & 0.154 & 0.092 & 0.116 &  0.072 & 0.080 & 0.044 & 0.062 & 0.068   \\
$2$& 0.4 & 0.426 & 0.424 & 0.416 & 0.256 & 0.294 &  0.186 & 0.186 & 0.074 & 0.148 & 0.124  \\
   & 0.6 & 0.722 & 0.736 & 0.732 & 0.508 & 0.590 &  0.354 & 0.354 & 0.128 & 0.312 & 0.316  \\
   & 0.8 & 0.924 & 0.922 & 0.932 & 0.792 & 0.852 &  0.592 & 0.586 & 0.230 & 0.516 & 0.514  \\
   & 1.0 & 0.988 & 0.988 & 0.986 & 0.902 & 0.964 &  0.754 & 0.766 & 0.388 & 0.716 & 0.716  \\ \hline
   & 0.0 & 0.052 & 0.060 & 0.040 & 0.038 & 0.062 &  0.020 & 0.030 & 0.048 & 0.020 & 0.050    \\
   & 0.2 & 0.642 & 0.656 & 0.600 & 0.318 & 0.272 &  0.680 & 0.690 & 0.194 & 0.400 & 0.290   \\
$6$& 0.4 & 0.998 & 1.000 & 0.996 & 0.946 & 0.896 &  1.000 & 1.000 & 0.708 & 0.970 & 0.900   \\
   & 0.6 & 1.000 & 1.000 & 1.000 & 1.000 & 1.000 &  1.000 & 1.000 & 0.966 & 1.000 & 1.000   \\
   & 0.8 & 1.000 & 1.000 & 1.000 & 1.000 & 1.000 &  1.000 & 1.000 & 1.000 & 1.000 & 1.000   \\
   & 1.0 & 1.000 & 1.000 & 1.000 & 1.000 & 1.000 &  1.000 & 1.000 & 1.000 & 1.000 & 1.000   \\
   \hline
\end{tabular} }
\end{table}

\textbf{Example 3.} We generate $X_1$ and $X_2$ independently from a uniform discrete distribution with 3 categories $\{-1,0,1\}$,
and let $Y=X_1+1.5|X_2|+ \varepsilon,$ where the random error $\varepsilon\sim N(0,1)$ or $t(1)$.
This simple example mimics a genetic association model where the SNPs are regressors and some continuous trait such as the body mass index is the response.
Note that the SNPs are categorical with three classes.
We apply the aforementioned methods to test the independence between $Y$ and $X_1$, $Y$ and $X_2$, respectively.
Table \ref{table.ex3} summarizes the empirical powers of each test based on 500 simulations at $\alpha=0.05$.
The DC test performs well when the random error is normal but the performance drops quickly when the extreme values are present.
The HHG test works well for testing the independence between $Y$ and $X_1$ but not for the pair of $Y$ and $X_2$.
The MV  test performs well in all settings.
It is also observed that the MV test based on the asymptotic null distribution (MV2) performs as well as the permutation-based MV test (MV1).

\begin{table}[ht]
\centering
\caption{\label{table.ex3} \it Empirical powers at $\alpha=0.05$ against the sample sizes in Example 3.}
\scalebox{1}{%
\begin{tabular}{cr|ccccc|ccccc}
  \hline
   &  &  \multicolumn{5}{c|}{$\varepsilon\sim N(0,1)$} & \multicolumn{5}{c}{$\varepsilon\sim t(1)$} \\ \hline
 $X_i$ & $n$ & MV1 & MV2 & DC & CS & HHG  & MV1 & MV2 & DC & CS & HHG  \\
  \hline
     &   30 & 0.814 & 0.832 & 0.744 & 0.648 & 0.754  & 0.370 & 0.370 & 0.238 & 0.274 & 0.384   \\
     &   60 & 0.992 & 0.994 & 0.976 & 0.952 & 0.962  & 0.668 & 0.682 & 0.394 & 0.554 & 0.656  \\
$X_1$&   90 & 1.000 & 1.000 & 0.998 & 0.998 & 0.994  & 0.870 & 0.878 & 0.594 & 0.782 & 0.872  \\
     &  120 & 1.000 & 1.000 & 1.000 & 1.000 & 1.000  & 0.964 & 0.974 & 0.704 & 0.926 & 0.970  \\
     &  150 & 1.000 & 1.000 & 1.000 & 1.000 & 1.000  & 0.988 & 0.986 & 0.800 & 0.966 & 0.988  \\ \hline
     &   30 & 0.616 & 0.624 & 0.640 & 0.388 & 0.316  & 0.286 & 0.276 & 0.190 & 0.208 & 0.142  \\
     &   60 & 0.916 & 0.932 & 0.926 & 0.784 & 0.684  & 0.550 & 0.552 & 0.358 & 0.418 & 0.348  \\
$X_2$&   90 & 0.990 & 0.992 & 0.992 & 0.958 & 0.910  & 0.728 & 0.744 & 0.498 & 0.636 & 0.512  \\
     &  120 & 1.000 & 1.000 & 1.000 & 0.990 & 0.966  & 0.850 & 0.864 & 0.636 & 0.786 & 0.670  \\
     &  150 & 1.000 & 1.000 & 1.000 & 1.000 & 0.996  & 0.944 & 0.944 & 0.720 & 0.890 & 0.800  \\
   \hline
\end{tabular}}
\end{table}

{
\textbf{Example 4.}
In this example, we follow Example 2 to generate data and let the number of classes  $R=20$
and the sample size $n=200$. Here, the signal vector $\mu=c(\mu_1,\mu_2,\ldots,\mu_{20}),$
where the values of $c$ vary from 0 to 1, each $\mu_j$ is randomly set to be one of $(1,2,3,4)$.
Note that the p-value for the MV test is computed using the approximated normal distribution
with mean $(R-1)/6$ and variance $(R-1)/45$ based on Theorem \ref{thm:main4} and others are based on the permutation tests.
It shows that the approximated normal null distribution of the MV statistic performs well for the large-$R$ case
and further supports  Theorem  \ref{thm:main4}.

\begin{table}[ht]
\centering
\caption{\label{table.ex4} \it Empirical powers at $\alpha=0.05$ against the signal strength for the large-$R$ case.}
\scalebox{1}{%
\begin{tabular}{c|cccc|cccc}
  \hline
     &  \multicolumn{4}{c|}{$X\sim N(0,1)$} & \multicolumn{4}{c}{$X\sim t(1)$} \\ \hline
  c &  MV & DC & CS & HHG   & MV & DC & CS & HHG   \\
  \hline
 0.0 &  0.052 & 0.056 & 0.042 & 0.068   &     0.050 & 0.042 & 0.042 & 0.056   \\
 0.2 &  0.366 & 0.382 & 0.182 & 0.166   &     0.118 & 0.110 & 0.062 & 0.068   \\
 0.4 &  0.974 & 0.982 & 0.284 & 0.722   &     0.502 & 0.152 & 0.130 & 0.126   \\
 0.6 &  1.000 & 1.000 & 0.818 & 0.966   &     0.896 & 0.414 & 0.296 & 0.246   \\
 0.8 &  1.000 & 1.000 & 0.990 & 1.000   &     0.982 & 0.672 & 0.542 & 0.432   \\
 1.0 &  1.000 & 1.000 & 1.000 & 1.000   &     1.000 & 0.892 & 0.814 & 0.578   \\
   \hline
\end{tabular}}
\end{table}

\subsection{A Real-Data Application}

\noindent
The colon cancer gene expression data set contains 62 tissue samples, which include 40 tumor biopsies from
colorectal tumors (labelled as ``negative") and 22 normal  biopsies from healthy parts of the colons (labelled as ``positive").
There are 2,000 genes which were selected out of more than 6,500 human genes  based on the confidence in the measured expression levels.
The data have been analyzed by \cite{Alon:1999} to reveal broad coherent patterns of correlated genes that suggested  a
high degree of organization underlying gene expression in these tissues. It is of interest to detect the significant genes
associated with the tissue syndrome.

We first applied the MV test to test for dependence between
genes and the tissue groups at the significance level $\alpha=0.05$. Since 2,000 hypotheses were simultaneously tested,
the Bonferroni correction was used to control the familywise error rate at 0.05. Thus, we
would test each individual hypothesis at the significance level $\alpha/2000=2.5\times 10^{-5}$.
The asymptotic null distribution in (\ref{nulldist}) was used to compute the p-value for each MV test
and 8 genes were identified as significance.
Figure \ref{figreal} displays the MV indices of all 2000 genes with the 8  significant genes.

\begin{figure}[!h]
\begin{center}
\begin{minipage}[c]{1\textwidth}
\centering
\includegraphics[scale=0.6]{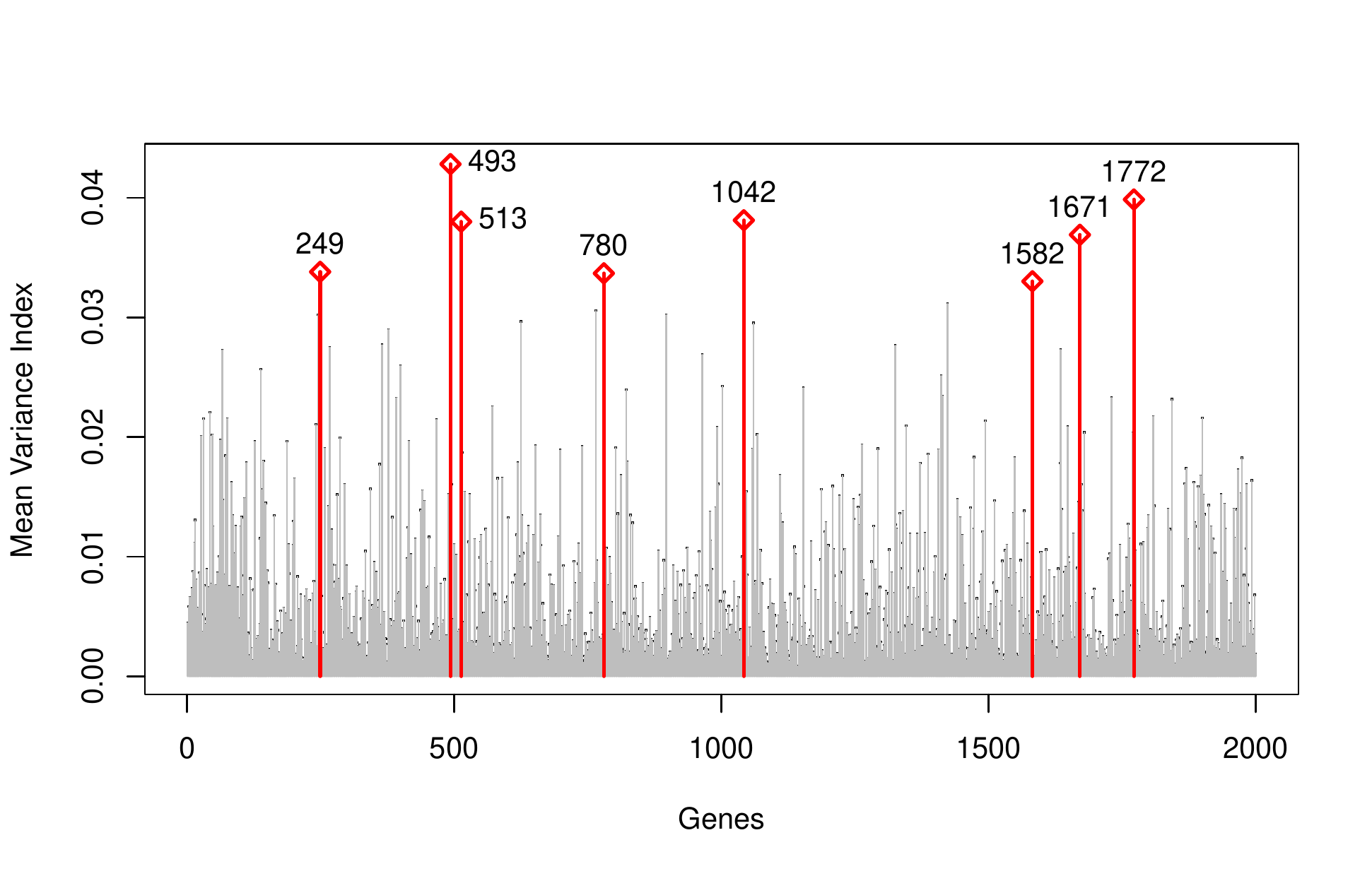}
\caption[]{\label{figreal} \small \it The MV indices of 2000 genes with the significant genes based on MV test.}
\end{minipage}
\end{center}
\end{figure}

Next, we applied the DC  test for the gene expression data. Note that
the smallest p-value obtained by the function \emph{dcov.test} using $K$ permutation times
is $1/(K+1)$. Thus, we chose $K=40000$ to make  the DC test applicable
to identify the significant genes.
For the DC  test, 12 genes were selected as significance.
Table \ref{tabreal} summarizes the computation time and the significant genes.
We conclude that it is very computationally efficient to conduct many simultaneous tests
using the explicit asymptotic null distribution compared with the permutation test,
since 2000 MV tests only took about 3 seconds.

To further check the significance of the selected genes, we randomly partitioned the data into
two parts: 80\% as the training data and the rest 20\% as the testing samples.  Then, the linear discriminant
analysis was applied to the training data based on the selected significant genes.
The classification accuracy (CA), i.e. the percentage of classifying test samples into the correct groups,
for the testing data was computed for each test and summarized in Table \ref{tabreal}.
All models had similar predication performance. However, the MV tests had the better prediction performance
based on the smaller set of significant genes. This result further demonstrated the MV test would be useful to test the significance of many genes simultaneously in high dimensional data analysis.

\begin{table}[ht]
\begin{center}
\caption{\label{tabreal} \it Comparison among different tests in the real-data application.}
\scalebox{1}{%
\begin{tabular}{c|ccccc}
  \hline
Tests &  Time(s) & CA &\# of Genes  & Indices of Significant Genes  \\  \hline
MV & 2.8   & 85.56\% & 8  & $\left\{{249~~493~~513~~780~~1042~~1582~~1671~~1772}\right\}$\\
DC & 603.4 & 85.01\% & 12  &
$\left\{\begin{array}{cccccc}
245 & 249  &267 & 377 & 493 & 765\\
822 &1042 &1325 &1423 &1582& 1772
 \end{array} \right\}$ \\
\hline
\end{tabular}
}
\end{center}
\end{table}

\section{Discussions}
\noindent
In this paper, we proposed the new distribution-free mean variance (MV) test of independence
 between a categorical random variable and a continuous one.
We derived an explicit form of its asymptotic null distribution, $\sum_{j=1}^{+\infty} {\chi_j^2(R-1)}/{\pi^2j^2}$, where $\chi_j^2(R-1), j=1,2,\ldots $, are independent $\chi^2$ random variables with $R-1$ degrees of freedom. It helps us to compute the empirical p-value efficiently in practice. It is also worth noting that this result does not depend on the distributions of two random variables $X$ and $Y$. Simulations and real data analysis showed its usefulness for detecting significant variables in high dimensional data.

Two extensions of the MV test can be considered. First, the MV test is also applicable in practice to test the independence between two continuous random variables by discretizing
 one continuous one into a categorical one. We can discretize a random variable $X$ using its percentiles  $\{\tau_1,\ldots, \tau_{K_n}\}$
 by defining  $X_{i}^* = k I(\tau_k\le X_{i} < \tau_{k+1}),$ where $I(\cdot)$ is an indicator function, $i=1,\ldots,n$, $k=1,\ldots,K_n$.
If $K_n$ is too large, then the sample size in each class is too small and the estimation of mean variance index is inaccurate.
By contrast, if $K_n$ is too small, then much information of the continuous variable may lose and the test power is unsatisfactory.
We can choose $K_n=O(n^{1/3})$ as \cite{Huang:Cui:2015} suggested.
In practice, we suggest to choose $K_n=[n/20]$, where $[a]$ means the integer part of $a$, so that the sample size in each category is around 20.
How to choose an optimal $K_n$ and the associated power performance will be left for the future research.
Second, another possible extension is to test the independent between a categorical response variable $Y$ and a random vector. Let $\bx=(X_1,\ldots, X_J)$ be a random vector with the dimensionality $J$. We can consider an aggregating approach to defining a multivariate MV between $Y$ and $\bx$ as
$MV(\bx|Y)=J^{-1}\sum_{1\le j\le J} MV(X_j|Y)$. The theoretical properties will be left for the future research.
\ \\

\section{Appendix: Proofs of Theorems}
\renewcommand{\theequation}{A.\arabic{equation}}
\renewcommand{\thesection}{A}
\setcounter{equation}{0}
To prove Theorem \ref{thm:main1}, we first need to define
\begin{eqnarray} \label{mvt1}
\widetilde{MV}(X| Y)  =\sum_{r=1}^R \frac 1{p_r}\int \left[\frac 1n\sum_{i=1}^n f_i(x,r)\right]^2 dF(x).
\end{eqnarray}
where $ f_i(x,r)=I\{X_i\leq x, Y_i=y_r\} - I\{X_i\le x\}
p_r-F(x)(I\{Y_i=y_r\} - p_r), $ for $i=1,\ldots,n$.
The following Lemma studies the difference between  $\widetilde{MV}(X| Y)$ and $\widehat{MV}(X| Y)$ under the null hypothesis of independence.
{\lem \label{lem31}
Under $H_0:$ $X$ and $Y$ are statistically independent, we have
\begin{eqnarray}
 \widehat{MV}(X| Y) - \widetilde{MV}(X| Y) = O_p\left(\frac {Rn^{-3/2}}{\underset{1\le r\le R}{\min} p_r}\right).
\end{eqnarray}
}
\noindent
\textbf{Proof of Lemma \ref{lem31}:}
First, we let
\begin{eqnarray}\label{mv1}
 \widetilde{MV}_1(X|Y) =\sum_{r=1}^R \frac 1{p_r}\int \left[\frac
1n\sum_{i=1}^n f_i(x,r)\right]^2 d\hat F(x),
\end{eqnarray}
where $ f_i(x,r)=I\{X_i\leq x, Y_i=y_r\} - I\{X_i\le x\}
p_r-F(x)(I\{Y_i=y_r\} - p_r), $ for $i=1,\ldots,n$.
Next, we consider the difference between $\widehat{MV}(X| Y)$ and $\widetilde{MV}_1(X| Y)$.
Note that
\begin{eqnarray*}
\widehat{MV}(X| Y)&=&\frac1n\sum_{r=1}^R\sum_{i=1}^n \hat p_r\left [\hat F_{r}(X_{i}) - \hat F(X_{i})\right]^2\\
&=&\sum_{r=1}^R \frac 1 {\hat p_r} \int\left[\hat F_{r}(x)\hat p_r- \hat F(x) \hat p_r\right]^2d\hat F(x)
\end{eqnarray*}
Thus, we have
\begin{eqnarray*}
&& \widehat{MV}(X| Y) -  \widetilde{MV}_1(X|Y) \\
&=& \sum_{r=1}^R \frac 1 {\hat p_r} \int\left[\hat F_{r}(x)\hat p_r- \hat F(x) \hat p_r\right]^2d\hat F(x)
-\sum_{r=1}^R \frac 1{p_r}\int \left[\frac 1n\sum_{i=1}^n f_i(x,r)\right]^2 d\hat F(x)\\
&=& \sum_{r=1}^R \left[\frac 1 {\hat p_r}- \frac 1{p_r}\right]
\int\left[\hat F_{r}(x)\hat p_r- \hat F(x) \hat p_r\right]^2d\hat F(x)\\
&+& \sum_{r=1}^R \frac 1{p_r}\int \left\{ \left[\hat F_{r}(x)\hat p_r- \hat F(x) \hat p_r\right]^2
- \left[\frac 1n\sum_{i=1}^n f_i(x,r)\right]^2 \right \}d\hat F(x)\\
&=:& I_{1n} + I_{2n} 
\end{eqnarray*}
We deal with the first term $I_{1n}$.
By the central limit theorem, we have $\hat p_r-p_r=O_p({n}^{-1/2})$.
Then,
$$  \frac 1 {\hat p_r}- \frac 1{p_r} = \frac{\hat p_r-p_r}{\hat p_rp_r}\le \frac {O_p({n}^{-1/2})}{\underset{1\le r\le R}{\min} p_r} .$$
Since
\begin{eqnarray*}&&
\hat F_{r}(x)\hat p_r- \hat F(x) \hat p_r  \\
&=& \frac 1n
\sum_{i=1}^n [I\{X_i<x, Y_i=y_r\} - F(x)p_r]  + F(x)p_r -\hat F(x)
(\hat p_r-p_r) - \hat F(x)p_r   \\ &=& \frac 1n \sum_{i=1}^n
[I\{X_i<x, Y_i=y_r\} - F(x)p_r] -\hat F(x) (\hat p_r-p_r) -[\hat
F(x) -F(x)]p_r,
\end{eqnarray*}
by the theory of empirical process, we have that
\bee \sup_x \left|\hat F_{r}(x)\hat p_r- \hat F(x) \hat p_r\right|
&\leq& \sup_x \left | \frac 1n \sum_{i=1}^n [I\{X_i<x, Y_i=y_r\} -
F(x)p_r \right| \\&+& |\hat p_r-p_r|
+ \sup_x\left|\hat F(x) -F(x)\right|\\
& =& O_p({n}^{-1/2}), \ene
where we note that
$E(I\{X_i<x, Y_i=y_r\})=E(I\{X_i<x\})E(I\{Y_i=y_r\})=F(x)p_r$
under the null hypothesis $H_0$.
It follows that
\begin{eqnarray} \nonumber
 I_{1n} &=& \sum_{r=1}^R \left[\frac 1 {\hat p_r}- \frac 1{p_r}\right]
\int\left[\hat F_{r}(x)\hat p_r- \hat F(x) \hat p_r\right]^2d\hat F(x)\\ \nonumber
&\leq&\sum_{r=1}^R \left | \frac 1 {\hat p_r}- \frac 1{p_r}\right |
\sup_x\left[\hat F_{r}(x)\hat p_r- \hat F(x) \hat p_r\right]^2\\ \label{i1n}
&=&  \frac {R}{\underset{1\le r\le R}{\min} p_r} O_p(n^{-3/2}).
\end{eqnarray}

Next, we deal with the second term $ I_{2n}$.
By the theory of empirical process, we have
\begin{eqnarray*}
&&\sup_x \left|\left[\hat F_{r}(x)\hat p_r- \hat F(x) \hat p_r\right]^2
- \left[\frac 1n\sum_{i=1}^n f_i(x,r)\right]^2 \right|  \\
&=& \sup_x \left|\hat F_{r}(x)\hat p_r- \hat F(x) \hat p_r -\frac 1n\sum_{i=1}^n f_i(x,r) \right|
\cdot \left| \hat F_{r}(x)\hat p_r- \hat F(x) \hat p_r   + \frac 1n\sum_{i=1}^n f_i(x,r)\right|\\
&=& {\sup_x\left|\hat F(x)-F(x)\right| \left|\hat p_r -p_r \right|}
\cdot\left\{ \sup_x\left|\hat F_{r}(x)\hat p_r- \hat F(x) \hat p_r \right| +
\sup_x\left|\frac 1n\sum_{i=1}^n f_i(x,r)\right| \right\} \\
&=& O_p(n^{-1/2})O_p(n^{-1/2})O_p(n^{-1/2})
= O_p(n^{-3/2}).
\end{eqnarray*}
where the second equality follows by
\begin{eqnarray*}
&&\hat F_{r}(x)\hat p_r- \hat F(x) \hat p_r   - \frac 1n\sum_{i=1}^n f_i(x,r)\\
&=& \left\{\frac 1n \sum_{i=1}^n
I\{X_i<x, Y_i=y_r\}  -\hat F(x) \hat p_r \right\}\\
&-&\left\{ \frac 1n \sum_{i=1}^n
I\{X_i<x, Y_i=y_r\}-\hat F(x)p_r-F(x)(\hat p_r-p_r) \right\}\\
&=& \left[\hat F(x)-F(x)\right] \left(\hat p_r -p_r \right).
\end{eqnarray*}
It follows that
\begin{eqnarray}\nonumber
I_{2n}&=&\sum_{r=1}^R \frac 1{p_r}\int \left\{ \left[\hat F_{r}(x)\hat p_r- \hat F(x) \hat p_r\right]^2
- \left[\frac 1n\sum_{i=1}^n f_i(x,r)\right]^2 \right \}d\hat F(x)\\ \label{i2n}
&=& \frac {R}{\underset{1\le r\le R}{\min} p_r} O_p(n^{-3/2}).
\end{eqnarray}
Thus, (\ref{i1n}) and (\ref{i2n}) together imply that
\begin{eqnarray}\label{diff1}
 \widehat{MV}(X| Y) -  \widetilde{MV}_1(X|Y) =\frac {R}{\underset{1\le r\le R}{\min} p_r} O_p(n^{-3/2}).
\end{eqnarray}

To complete the proof of  Lemma \ref{lem31}, it is sufficient to prove that
 the difference between $\widetilde{MV}_1(X|Y)$ and $\widetilde{MV}(X|Y)$ satisfies
that
\begin{eqnarray} \nonumber
\widetilde{MV}_1(X|Y) -  \widetilde{MV}(X|Y) &=&\sum_{r=1}^R \frac
1{p_r}\int \left[\frac 1n\sum_{i=1}^n f_i(x,r)\right]^2 d\left[\hat F(x) - F(x)\right] \\ \label{diff2}
&=& \frac {R}{\underset{1\le r\le R}{\min} p_r} O_p(n^{-3/2}).
\end{eqnarray}
It is enough to show
\begin{eqnarray}
 I_{3n}(r)=: \int \left[\frac 1n\sum_{i=1}^n f_i(x,r)\right]^2 d\left[\hat F(x) - F(x)\right] = O_p(n^{-3/2}).
 \end{eqnarray}
Without loss of
generality, we let $F(x)$ be the uniform distribution function,
since we can make the transformation $X'=F(X)$ for the continuous random variable $X$.
Therefore,
$$I_{3n}(r)= \frac 1n \sum_{j=1}^n\left[\frac 1n\sum_{i=1}^n
f_i(X_j,r)\right]^2 - \int_0^1 \left[\frac 1n\sum_{i=1}^n f_i(x,r)\right]^2 dx.
$$

For any $x,y\in (0,1)$, we can easily prove that
$$ E[f_i(x,r)f_j(y,r)]= (x\wedge y -xy)(p_r-p_r^2)I\{i=j\},$$
where $x\wedge y$ denotes the smaller value of $x$ and $y$.
\begin{eqnarray} \nonumber
 E[I^2_{3n}(r)]&=& E \left\{ \int_0^1 \left [ \frac 1n \sum_{j=1}^n
 [\bar f(X_j)^2 -  \bar f(x)^2]\right] dx \right\}^2 \\  \nonumber
&=& E\left\{ \int_0^1\int_0^1 \left[ \frac 1n \sum_{j=1}^n [ \bar f(X_j)^2 -  \bar f(x)^2]\right]
\left[ \frac 1n \sum_{j=1}^n [\bar f(X_j)^2 -  \bar f(y)^2]\right] dxdy \right\}\\ \nonumber
&=&  \int_0^1\int_0^1 E\left\{ [\bar f(X_1)^2-\bar f(x)^2 ][ \bar f(X_2)^2- \bar f(y)^2 ] \right\} dxdy
\end{eqnarray}
where $\bar f(x) =n^{-1}\sum_{i=1}^n f_i(x,r).$

Because $E(f_i(x,r))=0$ under $H_0$,
$E[f_i(x,r)f_j(x,r)f_k(y,r)f_l(y,r)]=0$ under $H_0$
if one of $\{i,j,k,l\}$ is different from the other three.
Then,  we can prove that
\begin{eqnarray*}\nonumber
E [\bar f(x)^2\bar f(y)^2] &=& \frac {1}{n^4}
\sum_{i,j}\sum_{k,l} E[f_i(x,r)f_j(x,r)f_k(y,r)f_l(y,r)] \\ \nonumber
&=& \frac 1{n^3} E[f_1(x,r)^2f_1(y,r)^2] + \frac
{n-1}{n^3}E[f_1^2(x,r)]E[f_2^2(y,r)] \\ \nonumber
&& +\frac {2(n-1)}{n^3}\left\{E[f_1(x,r)f_1(y,r)]\right\}^2 \\ \label{E1}
&=& O(n^{-3}) +  \frac{(p_r-p_r^2)^2}{n^2}[ xy(1-x)(1-y) + 2(x\wedge y -xy)^2].
\end{eqnarray*}
Similarly, we have
\begin{eqnarray*}\nonumber
 E [\bar f(X_1)^2\bar f(y)^2] &=& \frac {1}{n^4}
\sum_{i,j}\sum_{k,l} E[f_i(X_1,r)f_j(X_1,r)f_k(y,r)f_l(y,r)] \\ \nonumber
&=&
O(n^{-3}) +  \frac {(p_r-p_r^2)^2}{n^2} E[X_1y(1-X_1)(1-y) +
2(X_1\wedge y -X_1y)^2] \\ \label{E2}
&=& O(n^{-3}) +  \frac {(p_r-p_r^2)^2}{n^2} \int_0^1 [xy(1-x)(1-y) +
2(x\wedge y -xy)^2]dx.\\  \nonumber
 E [\bar f(x)^2\bar f(X_2)^2]
&=& \frac {1}{n^4} \sum_{i,j}\sum_{k,l}
E[f_i(x,r)f_j(x,r)f_k(X_2,r)f_l(X_2,r)]\\ \label{E3}
&=& O(n^{-3}) +  \frac {(p_r-p_r^2)^2}{n^2} \int_0^1 [xy(1-x)(1-y) +
2(x\wedge y -xy)^2]dy .\\ \nonumber
 E [\bar f(X_1)^2\bar f(X_2)^2]
&=& \frac {1}{n^4} \sum_{i,j}\sum_{k,l}
E[f_i(X_1,r)f_j(X_1,r)f_k(X_2,r)f_l(X_2,r)]\\ \label{E4}
&=& O(n^{-3}) +  \frac {(p_r-p_r^2)^2}{n^2} \int_0^1\int_0^1
[xy(1-x)(1-y) + 2(x\wedge y -xy)^2]dxdy .
\end{eqnarray*}
Therefore, we have
\begin{eqnarray} \nonumber
E[I^2_{3n}(r)]&=&
\int_0^1\int_0^1 E [\bar f(X_1)^2\bar f(X_2)^2] dxdy
-\int_0^1\int_0^1 E [\bar f(X_1)^2\bar f(y)^2] dxdy\\ \nonumber
&-&\int_0^1\int_0^1 E [\bar f(x)^2\bar f(X_2)^2] dxdy
+\int_0^1\int_0^1 E [\bar f(x)^2\bar f(y)^2] dxdy\\ \label{EI32}
&=& O(n^{-3}).
\end{eqnarray}
Because $E[I_{3n}(r)]=0$ for any $r$, we have $I_{3n}(r) = O_p(n^{-3/2}).$
 This completes the proof of Lemma \ref{lem31}.

\noindent
Lemma \ref{lem31} further implies that the difference between $T_n=n\widehat{MV}(X| Y)$ and $\widetilde{T}_n=n\widetilde{MV}(X| Y)$
is the order of $n^{-1/2}/\underset{1\le r\le R}{\min} p_r$ in probability. That is,
 $$T_n - \widetilde{T}_n = O_p\left(\frac {Rn^{-1/2}}{\underset{1\le r\le R}{\min} p_r}\right),$$
This lemma paves a road to derive the asymptotic null distribution of $T_n$ in Theorem \ref{thm:main1}.

\noindent \textbf{Proof of Theorem \ref{thm:main1}:} Based on the result of Lemma
\ref{lem31}, it is sufficient to prove that
\begin{eqnarray}\nonumber
n\widetilde{MV}(X| Y)  &=&\sum_{r=1}^R \frac 1{p_r}\int \left[\frac {1}{\sqrt n}\sum_{i=1}^n f_i(x,r)\right]^2 dF(x)\\  \label{eqn1}
&\overset{d}{\longrightarrow}&  \sum_{j=1}^{+\infty} \frac
{\chi_j^2(R-1)}{\pi^2j^2}
\end{eqnarray}

Denote $b_R=(\sqrt{p_1}, \sqrt{p_2}, \cdots, \sqrt{p_R})'$ and
$B=(b_1,b_2,\cdots, b_R)'$,
where $b_1,b_2,\ldots, b_{R-1} \in {\bf R}^R$
are $R-1$ unit and orthogonal vectors  such that
$B$ is an orthogonal matrix.
\bee {\bf
g}(x)&=&B\left(  \frac{ f(x,1)}{\sqrt{p_1}}, \frac {f(x,2)}{
\sqrt{p_2}}, \cdots, \frac {f(x,R)}{ \sqrt{p_R}} \right)',\\
\hat{\bf g}_i(x)&=&B\left(  \frac{ f_i(x,1)}{\sqrt{p_1}}, \frac
{f_i(x,2)}{ \sqrt{p_2}}, \cdots, \frac {f_i(x,R)}{ \sqrt{p_R}}
\right)', \ene where $f(x,r)=I\{X\leq x, Y=y_r\} - I\{X\le x\}
p_r-F(x)(I\{Y=y_r\} - p_r)$.

Let $ {\cal G} =\{ {\bf g}(x): x\in {\bf R}, r=1,2,\cdots, R\}$.
Since the graphical sets of $ I\{X\le x\}, I\{X\le x, Y=y_r\},
F(x)I\{Y=y_r\}$ and $F(x)$ form a Vapnik-Chervonenkis($VC$) class  respectively, then we
have that $\cal G$ forms a polynomial $VC$ class by the Lemma in
\cite{Pollard:1984}, and
\begin{eqnarray*}
 \left\{ \frac 1{\sqrt{n}}\sum_{i=1}^n \hat {\bf g}_i(x): x\in {\bf R} \right\}
\rightsquigarrow \left\{ Z(x)=(Z(x,1),Z(x,2), \cdots, Z(x,R))': x\in {\bf R}\right\}
\end{eqnarray*}
 by the Gaussian process convergence theorem in \cite{Pollard:1984} and \cite{Shorack:Wellner:1986},
 where $\rightsquigarrow$ denotes the convergence in distribution for any $x\in {\bf R}$, $\{ Z(x):  x\in {\bf R}\}$ is a Gaussian process with
$EZ(x)=0 $.

Let $C=(c_{rs})_{R\times R}$ be a $R\times R$ matrix with each element $c_{rs}$ defined by
$c_{rs}= E\left\{ \frac{ f(x,r)}{\sqrt{p_r}} \frac{ f(y,s)}{\sqrt{p_s}}\right\}$.
Since $f(x,r)=[I(X\le x)-F(x)][I(Y=y_r)-p_r]$ under $H_0$, then
\begin{eqnarray*}
&&c_{rs}= \frac{1}{\sqrt{p_rp_s}} E\left\{f(x,r)f(y,s)\right\}\\
&=&\frac{1}{\sqrt{p_rp_s}} E\Big\{[I(X\le x)-F(x)][I(Y=y_r)-p_r][I(X\le y)-F(y)][I(Y=y_s)-p_s]\Big\}\\
&=&\frac{1}{\sqrt{p_rp_s}} E\Big\{[I(X\le x)-F(x)][I(X\le y)-F(y)]\Big\}E\Big\{[I(Y=y_r)-p_r][I(Y=y_s)-p_s]\Big\}
\end{eqnarray*}
Note that $ E\Big\{[I(X\le x)-F(x)][I(X\le y)-F(y)]\Big\}=F(x)\wedge F(y)-F(x)F(y).$
If $s=r$, $E\Big\{[I(Y=y_r)-p_r][I(Y=y_s)-p_s]\Big\}=E\{[I(Y=y_r)-p_r]\}^2=p_r(1-p_r)$.
If $s\ne r$, $E\Big\{[I(Y=y_r)-p_r][I(Y=y_s)-p_s]\Big\}=-p_rp_s.$
Then
$$ C= [F(x)\wedge F(y)-F(x)F(y)][I_R- b_Rb_R'],$$
where $I_R$ denotes the $R\times R$ identity matrix.
Note that
$$B(I_R-b_Rb_R')B'=BB'-Bb_R(Bb_R)'=I_R-\mbox{diag}(0,\ldots,0,1)=\mbox{diag}(I_{R-1},0),$$
where  $I_{R-1}$ denotes the   $(R-1)\times (R-1)$ identity matrix.
Thus, we have
\begin{eqnarray} \nonumber
 E[Z(x,r)Z(y,s)] &=& E[{\bf g}(x) {\bf g}(y)']_{r,s}= (BCB')_{rs}\\ \nonumber
&=&[F(x)\wedge F(y)-F(x)F(y)] (B(I-b_Rb_R')B')_{rs}\\ \nonumber
&=&[F(x)\wedge F(y)-F(x)F(y)] (\mbox{diag}(I_{R-1},0))_{rs}\\ \label{zz}
 &=&\left\{ \begin{array}
{lll} &F(x)\wedge F(y)-F(x)F(y), &\ \ s=r=1,2,\cdots, R-1, \\
    & 0, &\ \ \mbox{otherwise}. \end{array}
  \right.
\end{eqnarray}

It implies that $Z(x,r)$ and $Z(y,s)$ are independent if $s\ne r$.
By applying the continuous mapping theorem, we have
\begin{eqnarray}
 \left\{ \left\|\frac 1{\sqrt{n}}\sum_{i=1}^n
\hat {\bf g}_i(x)\right\|^2: x\in {\bf R} \right \} \rightsquigarrow
\left\{ \|Z(x)\|^2: x\in {\bf R}\right\}.
\end{eqnarray}
Therefore,
\begin{eqnarray} \nonumber
 &&\sum_{r=1}^R \frac 1{p_r}\int\left [\frac 1{\sqrt{n}}\sum_{i=1}^n f_i(x,r)\right]^2 dF(x)\\ \label{eqn2}
 &=& \int \left\| \frac1{\sqrt{n}}\sum_{i=1}^n \hat {\bf g}_i(x) \right\|^2 dF(x)
 \overset{d}{\longrightarrow} \int \|Z(x)\|^2dF(x) \\ \label{eqn3}
 &=& \sum_{r=1}^{R-1}\int Z^2(x,r)dF(x) \overset{d}{=} \sum_{r=1}^{R-1} \sum_{j=1}^{+\infty}
\frac {\chi_{rj}^2(1)}{\pi^2j^2} \overset{d}{=} \sum_{j=1}^{+\infty} \frac
{\chi_{j}^2(R-1)}{\pi^2j^2} ,
\end{eqnarray}
where $\chi_{rj}^2(1)$'s denote the independently and identically distributed (i.i.d.) $\chi^2$ random variables with
$1$ degrees of freedom and  $\chi_j^2(R-1)$'s are i.i.d. $\chi^2$ random variables with
$R-1$ degrees of freedom,
the convergence in distribution $\overset{d}{\longrightarrow} $ follows the continuous mapping theorem,
the second equality sign is based on that $Z(x,r)$ and $Z(y,s)$ are independent if $s\ne r$ and the result (\ref{zz}),
the first $\overset{d}{=}$ is implied by Section 4.4 in \cite{Durbin:1973} or  Section 6.3.4 in \cite{Hajek:1999} .
This completes the proof of Theorem \ref{thm:main1}.

\noindent
\textbf{Proof of Theorem \ref{thm:main2}:}
Under the conditions assumed in Lemma 2.2 hold, we have,  under the alternative hypothesis $H_1$,
$\widehat{MV}(X| Y) \overset{p}{\rightarrow} MV(X|Y)>0$, as $n\rightarrow \infty$.
By Slutsky's theorem, we have
$T_n=n\widehat{MV}(X| Y)  \overset{p}{\longrightarrow} \infty, \mbox{~~as~~} n\to \infty.$
This completes the proof of Theorem \ref{thm:main2}.

\noindent
\textbf{Proof of Theorem \ref{thm:main3}:} For any $1\leq r\leq R$, we have $\int
(\hat F_r -\hat F)^2 d \hat F(x) = \int(F_r - F)^2dF(x) + o_p(1) $
and \bee  &&\int [(\hat F_r -\hat F)^2 -(F_r -F)^2]d \hat F(x) \\
&=&\int
[\hat F_r -\hat F-(F_r -F)][\hat F_r -\hat F +(F_r -F)] d \hat F(x)\\
&=& 2\int (F_r-F)[\hat F_r -\hat F-(F_r -F)]dF(x) + o_p(n^{-1/2})\\
&=&   \frac 2 {\hat p_r}\int (F_r-F)[\hat F_r\hat p_r -F_r \hat p_r
]dF(x) - 2\int (F_r-F)(\hat F -F)]dF(x) + o_p(n^{-1/2}) \\ &=& \frac
2 {p_r}\int (F_r-F)[\hat F_r\hat p_r -F_rp_r -F_r (\hat p_r -p_r)]dF(x)\\
&& - 2\int (F_r-F)(\hat F -F)]dF(x) + o_p(n^{-1/2})\\ &=& -\frac
2 {p_r} \int (F_r-F)F_rdF(x) (\hat p_r -p_r) + \frac 2{p_r} \int
(F_r-F)[\hat F_r\hat p_r -F_rp_r]dF(x) \\&& - 2\int (F_r-F)(\hat F
-F)]dF(x) + o_p(n^{-1/2}) . \ene
Thus, we have
\bee && \widehat {MV}(X|Y) - MV(X|Y) = \sum_{r=1}^R \left\{ \hat p_r \int
(\hat
F_r -\hat F)^2 d\hat F(x) -p_r \int (F_r -F)^2dF(x)\right\} \\
&=&\sum_{r=1}^R \left\{ (\hat p_r -p_r)\int (\hat F_r -\hat F)^2 d \hat
F(x) + p_r\int [(\hat F_r -\hat F)^2  -(F_r -F)^2]d \hat F(x) \right.  \\
&& \left.+ p_r \int (F_r -F)^2d (\hat F(x) -F(x)) \right\}\\
&=&\sum_{r=1}^R \left\{ (\hat p_r -p_r)\int (F_r -F)^2 dF(x)  + 2p_r\int (F_r-F) [\hat
F_r -\hat F-(F_r -F)]d F(x)\right.\\ &&
+ \left.p_r \int (F_r -F)^2d (\hat F(x) -F(x)) \right\} + o_p(n^{-1/2})\\
&=&\sum_{r=1}^R \left\{ \int (F^2 -F_r^2)dF(x)(\hat p_r -p_r)+2 \int
(F_r-F) [\hat F_r\hat p_r - F_rp_r]dF(x) \right. \\ && \left.-2p_r\int
(F_r-F)(\hat F -F)]dF(x)   + p_r \int (F_r -F)^2d (\hat F(x) -F(x))
\right\} + o_p(n^{-1/2}) \\
&=:& \frac 1n\sum_{i=1}^n \sum_{r=1}^RI_{4r}(X_i,Y_i)+
o_p(n^{-1/2}),
 \ene
where
{\color{black} \begin{eqnarray*}
I_{4r}(X,Y)&=:&\int (F^2 -F_r^2)dF(x)(I\{Y=y_r\}-p_r) \\
&+& 2\int (F_r-F)(I\{X<x,Y=y_r\} - F_rp_r)dF(x)\\ &-& 2p_r\int
(F_r-F)(I\{X<x\} -F(x))dF(x)\\ &+& p_r[(F_r(X) -F(X))^2- \int
(F_r(x) -F(x))^2dF(x)],
\end{eqnarray*}}
and $EI_{4r}(X_i,Y_i)=0$. Then by the Limit Central Theorem, we have
\begin{eqnarray*}
 \sqrt{n}[\widehat {MV}(X|Y) - MV(X|Y)] = \frac 1{\sqrt
n}\sum_{i=1}^n \sum_{r=1}^RI_{4r}(X_i,Y_i) + o_p(1) \overset{d}{\longrightarrow}
  N(0,\sigma^2).
\end{eqnarray*}
This completes the proof of Theorem \ref{thm:main3}.
\ \\

{\color{black}
\noindent \textbf{Proof of Theorem \ref{thm:main4}:}
First, we can prove the following three results.
\begin{eqnarray*}
(i). \ \ E[f_i(x,r)f_i(y,s)] &=& (x \bigwedge y -xy)(p_r\delta_{rs} -p_rp_s), \\
(ii). \ E[f_i^2(x,r)f_i^2(y,s)] &\leq &  C[p_r\delta_{rs}
+p_rp_s(p_r+p_s)],
\end{eqnarray*}
 for all $1\leq i \leq n, 1\leq r,s\leq R$,
where $C$ is a positive constant and $\delta_{rs}=1$ if $r=s$, $\delta_{rs}=0$ otherwise.

(iii). \begin{eqnarray*}
 \sum_{r,s,t,q=1}^R\frac 1 {p_rp_sp_tp_q}(p_r\delta_{rs}
-p_rp_s )( p_r\delta_{rt} -p_rp_t )(p_t\delta_{tq} -p_tp_q
)(p_s\delta_{sq} -p_sp_q ) = O(R).
\end{eqnarray*}

Then, with loss of generality, we assume that $X \sim
Unif(0,1)$, then $F(x)= x$ for $0 \leq x \leq 1$.
According to Lemma \ref{lem31}, we have
$$ T_n -\widetilde{T}_n = O\left( \frac {Rn^{-1/2}}{\underset{1\le r\le R}{\min} p_r} \right),$$
where $\widetilde{T}_n=n\widetilde{MV}(X| Y)$.
Then, under the condition $\sqrt{R} /\underset{1\le r\le R}{\min} p_r
=o(\sqrt{n}) $, we have that $T_n -\widetilde{T}_n =o(\sqrt R).$
Thus, it suffices to prove that
$$ \frac {\widetilde{T}_n - (R-1)/6}{\sqrt{(R-1)/45}} \overset{d}{\longrightarrow} N(0,1).
$$
Write
$$ \widetilde{T}_n =\frac 1n \sum_{r=1}^R \frac 1{p_r}\int \left[\sum_{i=1}^n f_i(x,r) \right]^2dx =:   J_{1n}+
J_{2n},
$$
where
$$ J_{1n}=\frac 1n \sum_{i=1}^n\sum_{r=1}^R \frac 1{p_r}\int
f_i^2(x,r)dx, \mbox{~~and~~}
J_{2n} =\frac 1n \sum_{i \ne j}^n\sum_{r=1}^R \frac 1{p_r}\int
f_i(x,r)f_j(x,r)dx.
$$
Note that $f_i(x,r)=(I(X_i\le x)-x)(I(Y_i=y_r)-p_r)$,
then
\begin{eqnarray*}
E(J_{1n}) &=& \sum_{r=1}^R \frac 1{p_r} E\left[\int
(I(X_1\le x)-x)^2(I(Y_1=y_r)-p_r)^2dx\right]\\
&=& \sum_{r=1}^R \frac 1{p_r} E\left[(I(Y_1=y_r)-p_r)^2\int
(I(X_1\le x)-2xI(X_1\le x)+x^2)dx\right]\\
&=& \sum_{r=1}^R \frac 1{p_r} E\left[(I(Y_1=y_r)-p_r)^2\times \frac16\right]\\
&=& \sum_{r=1}^R \frac 1{p_r} p_r(1-p_r)/6=(R-1)/6,
\end{eqnarray*}
and \bee  Var(J_{1n}) &=& \frac 1n
Var\left[\sum_{r=1}^R \frac 1{p_r}\int f_1^2(x,r)dx\right] \leq \frac 1n
E\left[\sum_{r=1}^R \frac 1{p_r}\int f_1^2(x,r)dx\right]^2 \\ &=& \frac 1n
\sum_{r,s}^R \frac 1{p_rp_s}\int\int E[f_1^2(x,r)f_1^2(y,s)]dxdy \\
&\leq& \frac Cn \sum_{r,s}^R \frac 1{p_rp_s}(p_r\delta_{rs} + p_rp_s
(p_r+p_s)) \\ &=& \frac Cn \left(\frac {R}{\min p_r} +R\right) = o(1).
 \ene
Next, we then only show that
$$\frac{J_{2n}}{\sqrt{(R-1)/45}} \overset{d}{\longrightarrow} N(0,1).
$$
Note that $E(J_{2n}) =0$ and \bee Var (J_{2n}) &=& E(J_{2n}^2)=
\frac 1{n^2}\sum_{i\ne j, k\ne l}^n \sum_{r,s}^R \frac 1{p_rp_s}
\int\int E[f_i(x,r)f_j(x,r)f_k(y,s)f_l(y,s)]dxdy \\ &=& \frac
{2n(n-1)}{n^2} \sum_{r,s}^R\frac 1{p_rp_s} \int\int
\{E[f_1(x,r)f_2(y,s)]\}^2dxdy  \\ &=& (1-\frac 1n) \frac {R-1}{45}.
\ene

Let ${\cal F}_i=\sigma\{(X_1,Y_1), \cdots, (X_i,Y_i)\}$. We also see
that \bee \frac {J_{2n}}{\sqrt {(R-1)/45}} =\frac {\sum_{i=2}^n
\big[ \frac 2 n \sum_{j=1}^{i-1}\sum_{r=1}^R \frac 1{p_r}\int
f_i(x,r)f_j(x,r)dx \big]}{\sqrt {(R-1)/45}}=:\sum_{i=2}^n Z_{ni}
\ene is the summation of a Martingale difference sequence with
$E(Z_{ni})=0$ and $Var[\sum_{i=2}^n Z_{ni}]=1-\frac 1n \to 1.$ We
need to prove  $\sum_{i=2}^n E(Z_{ni}^2 | {\cal F}_{i-1}) \overset{p}{\rightarrow} 1$.
Since $E[\sum_{i=2}^n E(Z_{ni}^2 | {\cal
F}_{i-1})] \to 1$ and \bee E(Z_{ni}^2|{\cal F}_{i-1}) &=& \frac
{45}{R-1} (\frac 2n)^2\sum_{j,k}^{i-1}\sum_{r,s}^R \frac
1{p_rp_s}\int\int E[f_i(x,r)f_i(y,s)]f_j(x,r)f_k(y,s)dxdy \\
&=&  \frac {45}{R-1} (\frac 2n)^2\sum_{j,k}^{i-1}\sum_{r,s}^R \frac
{p_r\delta_{rs} -p_rp_s}{p_rp_s}\int\int (x\bigwedge y
-xy)f_j(x,r)f_k(y,s)dxdy . \ene Thus, we have

\bee\sum_{i=2}^n E(Z_{ni}^2 | {\cal F}_{i-1}) &=&\frac {45}{R-1}
(\frac 2n)^2 \sum_{j=1}^{n-1} (n-j)\sum_{r,s}^R \frac
1{p_rp_s}\int\int E[f_i(x,r)f_i(y,s)]f_j(x,r)f_j(y,s)dxdy\\
&+& \frac {45}{R-1} (\frac 2n)^2 \sum_{j<k \leq n} (n-k)\sum_{r,s}^R
\frac 1{p_rp_s}\int\int E[f_i(x,r)f_i(y,s)]f_j(x,r)f_k(y,s)dxdy \\
&=:& J_{3n} + J_{4n}. \ene Since $ E(J_{3n}) \to 1$, $E(J_{4n})=0$
and $Var(J_{3n})\leq   \frac {CR^2\sum_{j=1}^{n-1}(n-j)^2 }
{(R-1)^2n^4} = O(1/n)$, then $J_{3n} \overset{p}{\rightarrow} 1$.

\bee Var(J_{4n})&=& (\frac {45}{R-1})^2 (\frac 2n)^4 \sum_{j<k, l<m}
(n-k)(n-m)\\ && E\Big (\sum_{r,s}^R \frac 1{p_rp_s}\int\int
E[f_i(x,r)f_i(y,s)]f_j(x,r)f_k(y,s)dxdy \\ && \sum_{r,s}^R \frac
1{p_rp_s} \int\int E[f_i(x,r)f_i(y,s)]f_l(x,r)f_m(y,s)dxdy\Big )
\\ &=&  (\frac {45}{R-1})^2 (\frac 2n)^4 \sum_{j<k, l<m}
(n-k)(n-m)O(R) = O(1/R).   \ene Therefore, $J_{4n} \overset{p}{\rightarrow} 0$.
On the other hand,  \bee \sum_{i=2}^n E(Z_{ni}^4)  &\leq&
\frac {C}{nR^2}
E\Big(\sum_{r=1}^R\frac 1{p_r}\int f_1(x,r)f_2(x,r)dx\Big)^4 \\
&\leq& \frac {C}{nR^2} \Big( \sum_{r,s}^R \frac 1{p_rp_s}\int\int
E[f_1^2(x,r)f^2(y,s)]dxdy\Big)^2 \\  &\leq& \frac {C}{nR^2}\Big(
\sum_{r,s}^R \frac 1{p_rp_s}[p_r\delta_{rs} +
p_rp_s(p_r+p_s)]\Big)^2 = O\left(\frac 1{n(\underset{1\le r\le R}{\min}{p_r})^2}\right)
=o(1/R) \ene By
the central limit theorem of the Martingale difference, we have
$$ \frac {\tilde {T}_n - (R-1)/6}{\sqrt{(R-1)/45}} \overset{d}{\longrightarrow} N(0,1),~~\mbox{as}~n\rightarrow \infty.
$$
This completes the proof of Theorem \ref{thm:main4}.
}

\noindent{\large\bf References}

\renewcommand{\baselinestretch}{1.00}
\baselineskip=14pt
\begin{description}

\newcommand{\enquote}[1]{``#1''}
\expandafter\ifx\csname
natexlab\endcsname\relax\def\natexlab#1{#1}\fi

\bibitem[{Agresti(2007)}]{Agresti:2007}
{\sc Agresti, A.} (2007). {\it An introduction to categorical data analysis,}
2nd ed. New York: John Wiley \& Sons.

\bibitem[Alon(1999)]{Alon:1999}
{\sc Alon, U., Barkai, N., Notterman, D. A., Gish, K., Ybarra, S., Mack, D.,} and {\sc Levine, A. J.} (1999).
{Broad patterns of gene expression revealed by clustering analysis of tumor and normal colon tissues probed by oligonucleotide arrays.}
\textit{Proc. Natl. Acad. Sci.} {\bf 96} 6745-6750.

\bibitem[Anderson and Darling(1952)]{AD:1952}
{\sc Anderson, T. W.,} and {\sc Darling, D. A.} (1952). {Asymptotic theory of certain `goodness-of-fit' criteria based on stochastic processes.}
\textit{Ann. Math. Statist.} {\bf23} 193-212.

\bibitem[{Bergsma and Dassios(2014)}]{Bergsma:Dassios:2014}
{\sc Bergsma, W.} and {\sc Dassios, A.} (2014). {A consistent test of independence based on a sign covariance related to Kendall's tau.}
\textit{Bernoulli} {\bf 20} 1006--1028.

\bibitem[{Cui, Li and Zhong(2015)}]{Cui:Li:Zhong:2014}
{\sc Cui, H., Li, R.} and {\sc Zhong, W.} (2015). {Model-free feature screening for ultrahigh dimensional discriminant analysis.}
 \textit{J. Amer. Statist. Assoc.}  \textbf{110} 630--641.

\bibitem[{Durbin(1973)}]{Durbin:1973}
{\sc Durbin, J.} (1973). {\it Distribution theory for tests based on the sample distribution function.} SIAM, Philadelphia, PA.

\bibitem[{Hajek, Sidak and Sen(1999)}]{Hajek:1999}
{\sc Hajek, J., Sidak, Z.} and {\sc Sen, P.} (1999). {\it Theory of rank tests,} 2nd ed. Academic Press, San Diego, CA.

\bibitem[{Heller, Heller and Corfine(2013)}]{HHG:2013}
{\sc Heller, R., Heller, Y.} and {\sc Corfine, M.} (2013). {A consistent multivariate test of association based
on ranks of distances.} \textit{Biometrika} {\bf100} 503--510.

\bibitem[{Hoeffding(1948)}]{Hoeffding:1948}
{\sc Hoeffding, W.} (1948). {A non-parametric test of independence.} \textit{Ann. Math. Statist.} {\bf19} 546--557.

\bibitem[{Huang and Cui(2015)}]{Huang:Cui:2015}
{\sc Huang, R.} and {\sc Cui, H.} (2015).
{Consistency of chi-squared test with varying number of classes.}
\textit{J. Sys. Sci. Complex.} {\bf28} 1--12.

\bibitem[{Kaufman(2014)}]{Kaufman:2014}
{\sc Kaufman, S.} and based in part on an earlier implementation by Ruth Heller and Yair
  Heller. (2014). HHG: Heller-Heller-Gorfine Tests of Independence. R package version
  1.4. \url{http://CRAN.R-project.org/package=HHG}


\bibitem[{Pollard(1984)}]{Pollard:1984}
{\sc Pollard, D.} (1984). {\it Convergence of stochastic processes.} Springer-Verlag, New York Inc.

\bibitem[{Pettitt(1976)}]{Pettitt:1976}
{\sc Pettitt, A. N.} (1976). {A two-sample Anderson-Darling rank statistic.} \textit{Biometrika.}
{\bf63} 161-168.

\bibitem[{Rizzo and Szekely(2014)}]{Rizzo:Szekely:2014}
{\sc  Rizzo, M. L.} and {\sc Sz\'{e}kely, G.  J.} (2014). energy: E-statistics (energy statistics).
  R package version 1.6.1. \url{http://CRAN.R-project.org/package=energy}

\bibitem[{Rosenblatt(1975)}]{Rosenblatt:1975}
{\sc Rosenblatt, M.} (1975). {A quadratic measure of deviation of two-dimensional density estimates and a test of independence.}
 \textit{Ann. Statist.}  {\bf3} 1--14.




\bibitem[{Shorack and Wellner(1986)}]{Shorack:Wellner:1986}
{\sc Shorack, G.} and {\sc Wellner J.} (1986).
{\it Empirical Processes with Applications in Statistics.}
Wiley, New York, Inc.

\bibitem[{Szekely,  Rizzo and  Bakirov(2007)}]{Szekely:Rizzo:Bakirov:2007}
{\sc Sz\'{e}kely, G.  J.,  Rizzo, M. L.} and  {\sc Bakirov, N. K.} (2007),
{Measuring and testing dependence by correlation of distances.}
 \textit{Ann. Statist.}  {\bf35}  2769--2794.

\bibitem[{Szekely and  Rizzo(2009)}]{Szekely:Rizzo:2009}
{\sc Sz\'{e}kely, G.  J.} and {\sc Rizzo, M. L.} (2009).
{Brownian distance covariance.}
 \textit{Ann. Appl. Statist.}  {\bf3}  1236--1265.


\end{description}

\end{document}